\documentclass[USenglish]{article}

\usepackage[%
english,    %
]{babel}

\usepackage[%
utf8        % Eingabekodierung auswaehlen
]{inputenc}

\usepackage[%
T1          % Mehr Buchstaben laden
]{fontenc}

\usepackage{lmodern}            % Schriftoptimierung fuer Standardschriftart

\usepackage{textcomp}

\usepackage{microtype}

\usepackage[                    %
affil-it                        % Institutsname kursiv
]{authblk}

\usepackage[%
format=plain,           % Unterschriften haengen NICHT am Bezeichner
labelsep=colon,               % Bezeichner wird durch Doppelpunkt abgetrennt
justification=justified,      % Blocksatz fuer Unterschrift
singlelinecheck=true,         % Kurze Bildunterschrift automatisch zentrieren
labelfont=bf                  % Bezeichner fett drucken
]{caption}

\usepackage{subcaption}

\usepackage[%
  style=authoryear,             % Bibliographiestil
  natbib=true,                  % Zitierstil
  safeinputenc=true,            % Korrigiere falsche Eingabekodierung
  maxbibnames=20,               % Maximale Anzahl Namen pro Bibliographieeintrag
  minbibnames=10,               % Minimale Anzahl Namen pro Bibliographieeintrag
  maxcitenames=2,               % Maximale Anzahl Namen pro Zitierung im Text
  hyperref=true,                % Verweis-Funktionalitaet aktivieren
  backref=false,                % Seitenzahl des Zitats in der Bibliographie
  isbn=false,                   % ISBN anzeigen
  url=false,                    % URL anzeigen
  doi=false,                    % DOI anzeigen
  eprint=false,                 % E-Print anzeigen
  firstinits=true,              % Initialen anzeigen
  uniquename=false,             % Eindeutige Namen sicherstellen
  uniquelist=false,             % Eindeutige Namen in Bibliographie
  sorting=nyt                   % Sortierreihenfolge
]{biblatex}

\bibliography{references.bib}        % Eingabedatei

\DeclareNameAlias{sortname}{last-first} % Reihenfolge der Namen

\usepackage{amsmath}
\usepackage{amsthm}
\usepackage{amsfonts}
\usepackage{amssymb}
\usepackage{amstext}

\newtheorem{remark}{Remark}[section]        % Bemerkung: Benennnung und Nummerierung
\usepackage{mathtools}

\usepackage{physics}

\usepackage{multirow}

\renewcommand{\epsilon}{\varepsilon}

\usepackage{siunitx}
\begin{document}

%%%%%%%%%%%%%%%%%%%%%%%%%%%%%%%%%%%%%%%%%%%%%%%%%%%%%%%%%%%%%%%%%%%%%%%%%%%%%%%%
%% Titelseite
\microtypesetup{protrusion=false}
\title{A model for warm clouds with implicit droplet activation, avoiding saturation adjustment}

\author[1]{Nikolas Porz}
\author[2]{Martin Hanke}
\author[3]{Manuel Baumgartner}
\author[1]{Peter Spichtinger}

\affil[1]{                      %
  Institute for Atmospheric Physics, Johannes Gutenberg University, Mainz, Germany
}
\affil[2]{                      %
  Institute of Mathematics, Johannes Gutenberg University, Mainz, Germany
}
\affil[3]{Data Center, Johannes Gutenberg University, Mainz, Germany}

\date{\today}
\maketitle
\microtypesetup{protrusion=true}
%%%%%%%%%%%%%%%%%%%%%%%%%%%%%%%%%%%%%%%%%%%%%%%%%%%%%%%%%%%%%%%%%%%%%%%%%%%%%%%%

%%%%%%%%%%%%%%%%%%%%%%%%%%%%%%%%%%%%%%%%%%%%%%%%%%%%%%%%%%%%%%%%%%%%%%%%%%%%%%%% 
% Zusammenfassung
\begin{abstract}
The representation of cloud processes in weather and climate models 
is crucial for their feedback on atmospheric flows. Since there is no general
macroscopic theory of clouds, the parameterization of clouds in corresponding
simulation software depends crucially on the underlying modeling assumptions.
In this study we present a new model of intermediate complexity 
(a one-and-a-half moment scheme) for warm clouds, which is derived from 
physical principles. 
Our model consists of a system of differential-algebraic equations
which allows for supersaturation and comprises intrinsic automated
droplet activation due to a coupling of the droplet mass- and number 
concentrations tailored to this problem.
For the numerical 
solution of this system we recommend a semi-implicit integration scheme, 
with efficient solvers for the implicit parts.
The new model shows encouraging numerical results when compared with 
alternative cloud parameterizations, and it is well suited to
investigate model uncertainties and to quantify predictability of weather
events in moist atmospheric regimes.
\end{abstract}
%%%%%%%%%%%%%%%%%%%%%%%%%%%%%%%%%%%%%%%%%%%%%%%%%%%%%%%%%%%%%%%%%%%%%%%%%%%%%%%% 

%%%%%%%%%%%%%%%%%%%%%%%%%%%%%%%%%%%%%%%%%%%%%%%%%%%%%%%%%%%%%%%%%%%%%%%%%%%%%%%%
%%%%%%%%%%%%%%%%%%%%%%%%%%%%%%%%%%%%%%%%%%%%%%%%%%%%%%%%%%%%%%%%%%%%%%%%%%%%%%%%
%%%%%%%%%%%%%%%%%%%%%%%%%%%%%%%%%%%%%%%%%%%%%%%%%%%%%%%%%%%%%%%%%%%%%%%%%%%%%%%%
%%%%%%%%%%%%%%%%%%%%%%%%%%%%%%%%%%%%%%%%%%%%%%%%%%%%%%%%%%%%%%%%%%%%%%%%%%%%%%%%

\section{Introduction} %/Motivation}
\label{sec:introduction}

Clouds are important components in the Earth-Atmosphere system. They
influence the hydrological cycle via precipitation formation, the
organization of weather phenomena (convection etc.), and the energy
budget by their interaction with radiation. It is well known that clouds
constitute a major source for forecast errors for weather prediction,
or more precisely, they influence the predictability of moist
atmospheric flows in a crucial way. %\todo[inline]{noch referenzen finden}
This is mostly due to the fact that diabatic processes
(e.g. latent heating or mixing) 
serve as large energy sources and sinks
\citep[e.g.][]{joos_wernli2012,igel_vandenheever2014}.  

% PS eingefügter text
% The predictability of atmospheric flows is crucially affected by the
% representation of microphysical cloud processes in models.
% PS 

The representation of clouds in models for weather forecast and
climate prediction is an important and challenging task. Cloud
processes take place on a variety of scales and interact with other
processes (e.g. atmospheric flows) in a truly multiscale fashion.
Consisting of a myriad of small water particles which evolve in space
and time, a rigorous simulation based on fundamental physical
principles is way beyond current computing capacities.  Standard
implementations therefore resort to certain parameterizations of the
cloud system; depending on the level of sophistication there exist (i)
single moment schemes, which only keep track of the spatial water mass
concentration for certain types of particles
\citep[e.g.][]{kessler1969,lin_etal1983,cosmo_doc_physical_parameterization},
(ii) double moment schemes, which also monitor the number
concentrations of these particles
\citep[e.g.][]{morrison_etal2005,seifert_beheng2006}, and (iii)
statistical models, which assume a statistical distribution of the
particles over an admissible range of mass values, and then evolve the
distribution function in time (and space), which leads to
Boltzmann-type evolution equations \citep[see,
e.g.,][]{beheng2010,khvorostyanov1995,khain_etal2000}.

While the latter ansatz seems to be very attractive, there are several
problems associated with it. First, at least to our knowledge, no
consistent treatment of all cloud processes has yet been achieved with
such a setting, although several attemps have been made in the
literature: usually, these formulations concentrate on the collision
terms but omit other important processes like, for example, particle
formation \citep[e.g.][]{beheng2010}.  Second, it is often assumed
that the type of the distribution does not change in time, but this
assumption is violated for almost all important cloud processes
\citep[e.g.][]{gierens_bretl2009}.  Third, even for an incomplete
version of the corresponding evolution equations, their numerical
treatment is quite difficult and expensive. Finally, measurements of
the mass distribution of cloud particles are very difficult to
realize, so that real data are lacking for model calibration.

On the other hand, real measurements are available for number and mass
concentrations of the water droplets, i.e., for the corresponding
variables of a two-moment scheme \citep[e.g.][]{wendisch_etal2016}. In
the statistical ansatz those correspond to certain moments of the
distributions.  We found that when focussing only on these number and
mass concentrations, then a methodology which is well-known from
chemical reaction networks and population dynamics leads essentially
to the same dynamical system as when starting from a statistical
description of the ensemble.

In the formulation of models based on averaged quantities like number
and mass concentrations, the following problems need to be addressed:

\begin{enumerate}
\item Collision terms:\\
  The formulation of collision terms is not straightforward, since the
  details and the evolution of the underlying size distribution must 
  be mimicked in an adequate way. The standard separation
  of non-falling cloud droplets and falling rain drops due to
  \citet{kessler1969} leads to artificial processes called autoconversion
  and accretion, which must be parameterized in a meaningful way. 
\item Particle generation:\\
  The formation of cloud droplets is based on the activation of
  cloud condensation nuclei (CCN) in the atmosphere
  \citep{koehler1936}. For a proper treatment of this process,
  aerosols and their chemical and physical properties must be taken
  into account, which is difficult to model appropriately; it would
  require the extension of the model to also include aerosol physics.
\item Growth/evaporation of liquid droplets:\\
  The representation of condensation processes can change the
  distribution of latent heating and thus influence the evolution of
  convective systems \citep[e.g.][]{marinescu_etal2017}.
% In particular,
% the use of saturation adjustment has a large effect on the vertical
% evolution of convection, since the latent heating is overestimated
% leading to errors in buoyancy
% \citep[][]{grabowski_morrison2016,grabowski_morrison2017}; these
% effects turned out to be quite significant.
  Small cloud droplets grow exclusively by diffusion in the
  supersaturated regime, which is very fast for relevant temperatures;
  this leads to stiff differential equations and numerical
  instabilities.  Many models therefore consider a technique known as
  saturation adjustment \citep[e.g.][]{kogan_martin1994}, assuming
  water clouds at water saturation.  However, this approach has a
  large effect on the vertical evolution of convection, since the
  latent heating is overestimated leading to significant errors in
  cloud buoyancy \citep[cf.][]{grabowski2015,grabowski_morrison2016,
    grabowski_morrison2017}.
  % physical inconsistencies and errors in the predicted buoyancy
  % \citep[cf.][]{grabowski2015,grabowski_morrison2016}.
%% Referee: quantify
\end{enumerate}

In addition to a consistent formulation of the
processes and the model, we also need to discuss appropriate numerical
schemes for solving the equations in an adequate way. Finally, we
have to make sure that meaningful solutions exist.

Our motivation for the development of yet another cloud model based on
bulk variables is as follows. To improve predictability of moist
atmospheric flows, the adequate representation of clouds and their
processes is an important issue. These investigations are pushed
forward from the point of view of atmospheric dynamics and weather
forecasts in connection with operational weather forecast models, as,
e.g., COSMO or, more recent ICON, driven by the German Weather
Service, or the IFS model used at the European Centre for Medium-Range
Weather Forecasts.  These operational models use very simple schemes
for representing clouds. Usually, single moment schemes are used,
which predict mass concentrations of cloud and rain water, only. It is
well known that especially collision processes cannot be represented
in a sufficient way for single moment schemes, 
%% MHB: reformulated 
which impedes an accurate prediction of the formation of rain.
In addition,
%the generation of cloud droplets (i.e., activation of cloud
%condensation nuclei to droplets) 
CCN activation cannot be treated well
% in a physically consistent way 
in such simple models; mostly it is assumed that clouds
exist at thermodynamic equilibrium and the number of cloud droplets is
prescribed. This gives rise to additional uncertainties and
forecast errors.

On the other hand, there are complex cloud models
available for research purposes with sophisticated schemes for the
treatment of collision processes and particle generation \citep[see,
e.g.,][]{seifert_beheng2006,morrison_grabowski2008,morrison_milbrandt2015}.
%% Referee: erlaeutern!
However, it turns out that the investigation of such complex models
and their impact on dynamics is very complicated; due to the complex
and sometimes discrete formulation of the processes the estimation of
the impact of these processes on dynamics is almost impossible. 

As another important issue, the coefficients of
many process rates in the governing equations are not well
determined from first principles. Often, these rates are only known to belong
to a certain range, but their exact values must be estimated by comparison
with measurements or reference models. For the quantification of
predictability of moist flows including clouds, we have to use inverse methods
to investigate the uncertainties in these parameters first.
For this reason it is desirable to have a model of only intermediate 
complexity.

In this study we develop 
%a consistent and simple model, which can be used
such a model, well suited for a mathematical analysis of the associated
processes and for the estimation of its parameters.
To be precise, we develop a one-and-a-half moment scheme,
i.e. a set of differential equations for mass concentration of cloud
droplets and mass and number concentrations for rain drops. 
We solve the activation problem by introducing a functional relation between 
cloud droplet number and mass concentrations. Finally, cloud droplet growth
and evaporation are treated using the diffusional growth equation, i.e., no
saturation adjustment is used. For the proper treatment of these
equations, we present a consistent 
sophisticated numerical scheme.

The paper is structured as follows: In the next section we describe
the model including the relevant processes, the process rates, and
their representation in mathematical terms. 
Subsequently, in Section~\ref{sec:numericalscheme}, we develop our numerical 
scheme, and we present some numerical results in 
Section~\ref{sec:numtest}. We end the study with a short summary and
some conclusions in Section~\ref{sec:conclusion}.

\section{Model description}
\label{sec:model_description}

Our model can be used as a box model for a single volume parcel or as
a model of a vertical column of air, which is transported in a
Lagrangian way (i.e., along a 
%% MHB: inserted ``given''
given trajectory). For the latter case we
denote the vertical spatial coordinate by $z$.
At the moment, the model is not coupled to any equations for 
atmospheric flows (i.e., Navier-Stokes equations, or relevant
approximations), although this may be carried out in future work. A
simpler version of our model was already implemented within a flow solver
\citep{lukacova_etal2017}. 

We restrict our model to so-called warm clouds or liquid clouds, which
commonly occur in the temperature regime
$\SI{250}{K}<T<\SI{310}{K}$, i.e., ice processes have not been included into
the model. An extension into this direction is planned, but is beyond
the scope of this work.% \textbf{no ice phase --> later}

\subsection{Basic assumptions}

%Clouds consist of huge number of water particles. In our case we
%We restrict our model to so-called warm clouds or liquid clouds, which
%commonly occur in the temperature regime
%$\SI{250}{K}<T<\SI{310}{K}$. 
In warm clouds one can distinguish two water phases, namely water
vapour and liquid droplets of various sizes.  The droplets can
interact with each other and also with water vapour, depending on the
thermodynamic conditions (i.e., temperature, pressure, and water
vapour concentration).  Measurements indicate two well separated modes
in the size/mass distribution of liquid water particles \citep{warner1969}.
%Since a particle based
%description of these interactions would lead to a huge system of
%equations, the ensemble has to be treated in a statistical way.  This
%can be done by introducing statistical distributions for mass or size
%as functions of space and time, and the evolution of these statistical
%distributions then leads to Boltzmann-type evolution equations
Therefore, we distinguish two species of water particles, namely cloud
droplets (index $c$) and rain drops (index $r$), and as we have already
said, we keep track of the bulk variables
mass concentration $q_x$ and number concentration $n_x$ for each 
water particle species $x$, rather than their statistical distributions.
To be precise, % our model might be considered a one-and-a-half moment scheme, 
% because
we only evolve number and mass concentrations for rain drops 
independently, whereas we couple the number concentration $n_c$ of 
cloud droplets to their mass concentration $q_c$ via a sophisticated functional
relation, cf.~(\ref{eq:nc_relation}).

As it is standard in cloud physics, number and mass concentrations are 
given in units per mass dry air,
i.e.,  $[n_x]=\si{\per\kilogram}$ and $[q_x]=\si{\kilogram\per\kilogram}$.  For
simplification, we assume that cloud and rain drops are spherical,
so that the mass $m$ of a water droplet (i.e., $m_c$ or $m_r$) is given by the 
radius $r$ of this droplet via
\begin{equation}
  \label{eq:mass_radius}
  m=\frac{4}{3}\pi r^3 \rho_l,
\end{equation}
where $\rho_l$ denotes the (volumetric) density of liquid water. We
also employ temperature $T$, pressure $p$, and water vapour
concentration $q_v$ as thermodynamical variables. To a very good
approximation we can assume air (index $a$) and water vapour (index
$v$) as ideal gases, using the ideal gas law
%% MHB: changed index k to x (ten times!)
\begin{equation*}
  \label{eq:ideal_gas}
  \begin{aligned}
    p_x V = M_xR_xT& &\Longleftrightarrow& &p_x=\rho_x R_x T
  \end{aligned}
\end{equation*}
with $M_x$ the mass, and $R_x=R^*/M_{\text{mol},x}$ the specific gas constant 
for the substance $x\in\{a,v\}$, given the universal gas constant $R^*$ and
the respective molar mass $M_{\text{mol},x}$. Generally, Dalton's law
is applied, i.e., the total pressure is assumed to satisfy
$p=p_a+p_v$. Since $p_v\ll p_a$ and $\rho_v=q_v\rho_a\ll\rho_a$ we
usually use the approximation $p\approx p_a$ and
$\rho=\rho_v+\rho_a\approx\rho_a$. We thus have
\begin{equation}
  \label{eq:gaslaw}
  \begin{aligned}
    \rho=\frac{p}{R_aT},& &R_a=\SI{287.05}{\joule\per\kilogram\per\kelvin},
  \end{aligned}
\end{equation}
to a high level of accuracy.

The thermodynamic equilibrium between water vapour and
liquid water is determined by the Clausius-Clapeyron equation,
describing the saturation vapour pressure $p_s(T)$.  For the latter we
use the approximation provided in Section~\ref{appendixB:constants}
due to \citet{murphy_koop2005}, compare \eqref{eq:appendixB_pliq}. 
The saturation vapour concentration for water vapour can be approximated by
\begin{equation}
  \label{eq:vs}
  \begin{aligned}
    q_{vs}(p,T)\approx \epsilon\,\frac{p_s(T)}{p},& &\epsilon=\frac{M_{\text{mol},v}}{M_{\text{mol},a}}\approx 0.622.
  \end{aligned}
\end{equation}
The latent heat of the phase
transition between vapour and liquid water is set to the constant
value $L=\SI{2.53e6}{\joule\per\kilogram}$.

%% Referee: Give references for the following list...
For our description of warm clouds we make the following assumptions:
\begin{enumerate}
\item We distinguish our two liquid species according to their
  particle sizes: cloud droplets are small with a diameter below
  $\SI{50}{\micro\metre}$, in general, while rain drops are much
  larger. 
\item Since cloud droplets are small, their sedimentation velocity is
%% MHB: slightly changed the formulation
  negligible. Rain drops, on the other hand, are large enough to fall, and
  they have a terminal fall velocity $v_t$ depending on their size, as
  can be derived from theory and measurements
  \citep{pruppacher_klett2010}. This distinction has been introduced by
  \citet{kessler1969} for the first time.
\item Only cloud droplets can be formed out of the gas phase,
  i.e., water droplets grow from activated aerosols due to Köhler
  theory \citep{koehler1936,arabas_shima2017}.
\item Cloud droplets and rain drops can grow and evaporate due to diffusion
  of water vapour, but we neglect diffusional growth of rain
  drops, since it is very slow \citep{devenish_etal2012}; this
  assumption is used in many models \citep[e.g.][]{lin_etal1983}.
\item Rain drops form and grow by collision of/with cloud droplets;
  this is the major pathway for the growth of large water particles
  \citep[][]{khain_etal2000}. 
\end{enumerate}

\begin{figure}%[h!]
  \centering
  \includegraphics[height=0.505\linewidth]{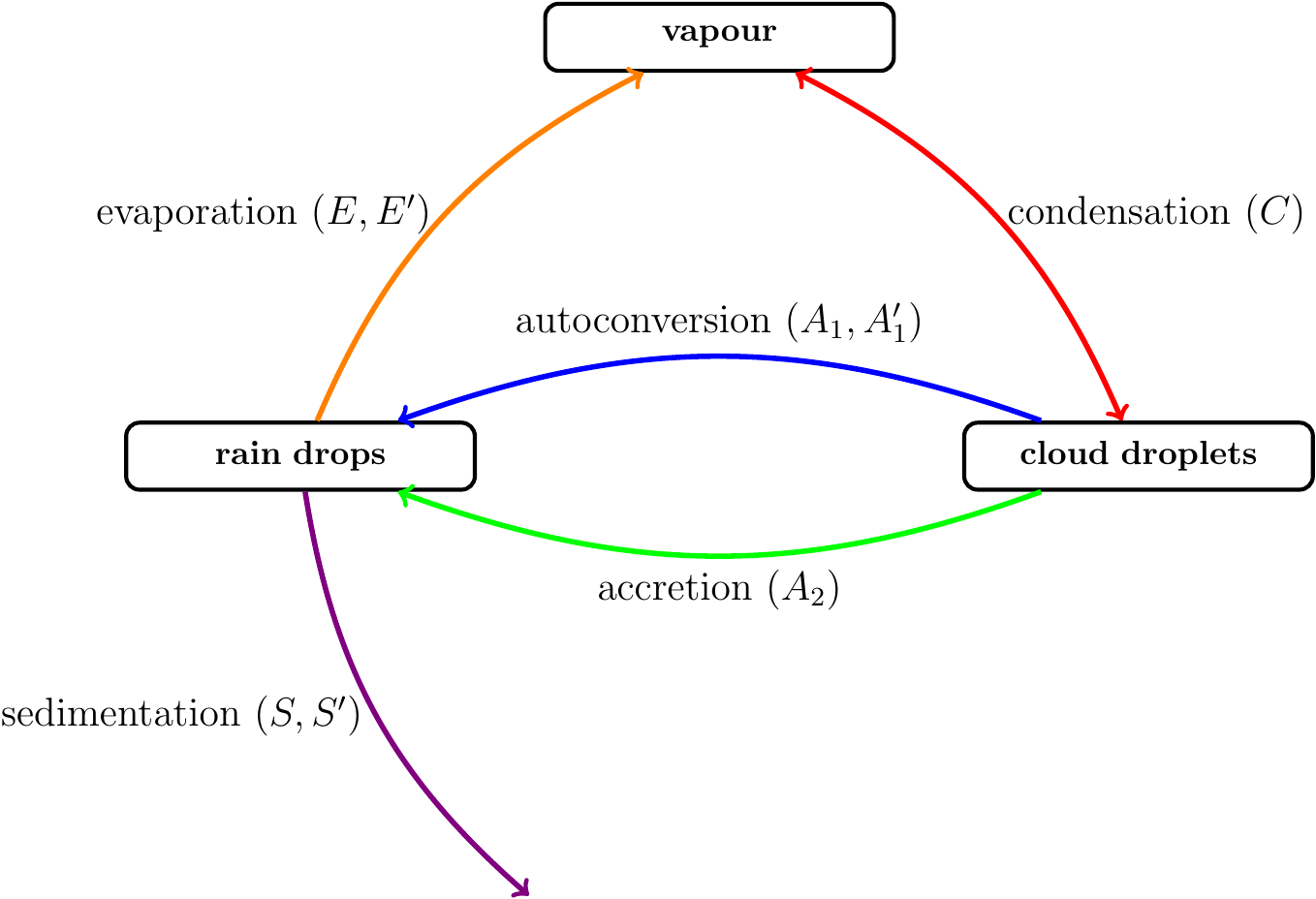}
%  % {model_figure.pdf} 
%  \includegraphics[width=0.75\linewidth]{ProzessDiagramm.pdf}
  % {model_figure.pdf} 
  \caption{Processes and interactions between the water species (vapour, cloud droplets and rain drops) in the model.\label{fig:processes}}
\end{figure}

%% MHB: changed ``consistent'' to ``corresponding'' and ``cloud'' to ``cloud system''
For the formulation of a corresponding system of equations for the time
evolution of the cloud system we thus consider the following processes
(see Figure~\ref{fig:processes}):

%% MHB: added the `primed' processes
\begin{itemize}
\item Formation of cloud droplets due to condensation (process $C$).
\item Growth/evaporation of cloud and rain particles due to diffusion
  (processes $C$, $E$, $E'$).
\item Collision of particles for forming rain drops
 (processes $A_1$, $A_1'$, $A_2$).
\item Sedimentation of rain drops (processes $S$, $S'$).
\end{itemize}

%The processes of the model and their interaction with the different
%water species are represented in Figure~\ref{fig:processes} 

As a general rule we first investigate rates on a single particle
basis, if possible. In a second step we extend this ansatz to derive
corresponding rates for the bulk variables mass and number
concentrations, respectively. This then yields differential equations 
for the cloud variables $q_c,q_r,n_r$, and the
thermodynamic variables $q_v,p,T$, which are coupled to additional 
algebraic state equations for $n_c$ and $\rho$ for closing the system.

\subsubsection{Terminal velocity of water particles}

A water droplet of spherical shape is accelerated by gravity. On the
other hand, friction of air is changing momentum in the opposite direction. 
Eventually, the balance of forces leads to a constant velocity of the
particle, the so called terminal velocity $v_t$. There are different
descriptions of $v_t$ in the pertinent literature;
as our gold standard we quote from
\citep[][their eq.~(4)]{seifert_etal2014} the formula 
\begin{equation}
  \label{eq:vt_ref}
  v_t(r)=\alpha_r-\beta_r\,\exp(-\gamma_r2r)
\end{equation}
with
\begin{equation*}
  \begin{aligned}
    \alpha_r=\SI{9.292}{\meter\per\second},& &\beta_r=\SI{9.623}{\meter\per\second},& &\gamma_r=\SI{6.222e2}{\per\meter}
  \end{aligned}
\end{equation*}
for larger drops with radius $r>\SI{50}{\micro\meter}$. %\SI{0.05}{mm}$.
A simpler approximation from \citet{seifert_beheng2006} uses the power
law ansatz
\begin{equation}
  \label{eq:vt_sb06}
  \begin{aligned}
    v_t(m)= \alpha' \,m^\beta
%    \left(\frac{\rho_*}{\rho}\right)^{\frac{1}{2}},
    & &
  \alpha'=\SI{159}{\meter\per\second\per\raiseto{\beta}\kilogram},& &\beta=\frac{4}{15}\,,
\end{aligned}
\end{equation}
depending on the drop mass $m$, connected to the radius 
via eq.~\eqref{eq:mass_radius}.
Note, that this approximation was formulated originally by
\citet{liu_orville1969}.  
These approximations of $v_t$ provide reference values of the terminal
velocity, which correspond to the density
\[
  \rho \,=\, \rho_*\,=\, \SI{1.225}{\kilogram\per\cubic\meter}
\]
of dry air at normal pressure
$p_*=\SI{101325}{\pascal}$ and temperature $T_*=\SI{288}{\kelvin}$.
For other densities they have to be adapted, using the ansatz
$(\rho_*/\rho)^c$ as discussed in \citet[][their Appendix
A]{naumann_seifert2015}, with the exponent
$c=\frac{1}{2}$ for large rain drops.

\begin{figure}%[t!]
  \centering
  \includegraphics[height=0.505\linewidth]{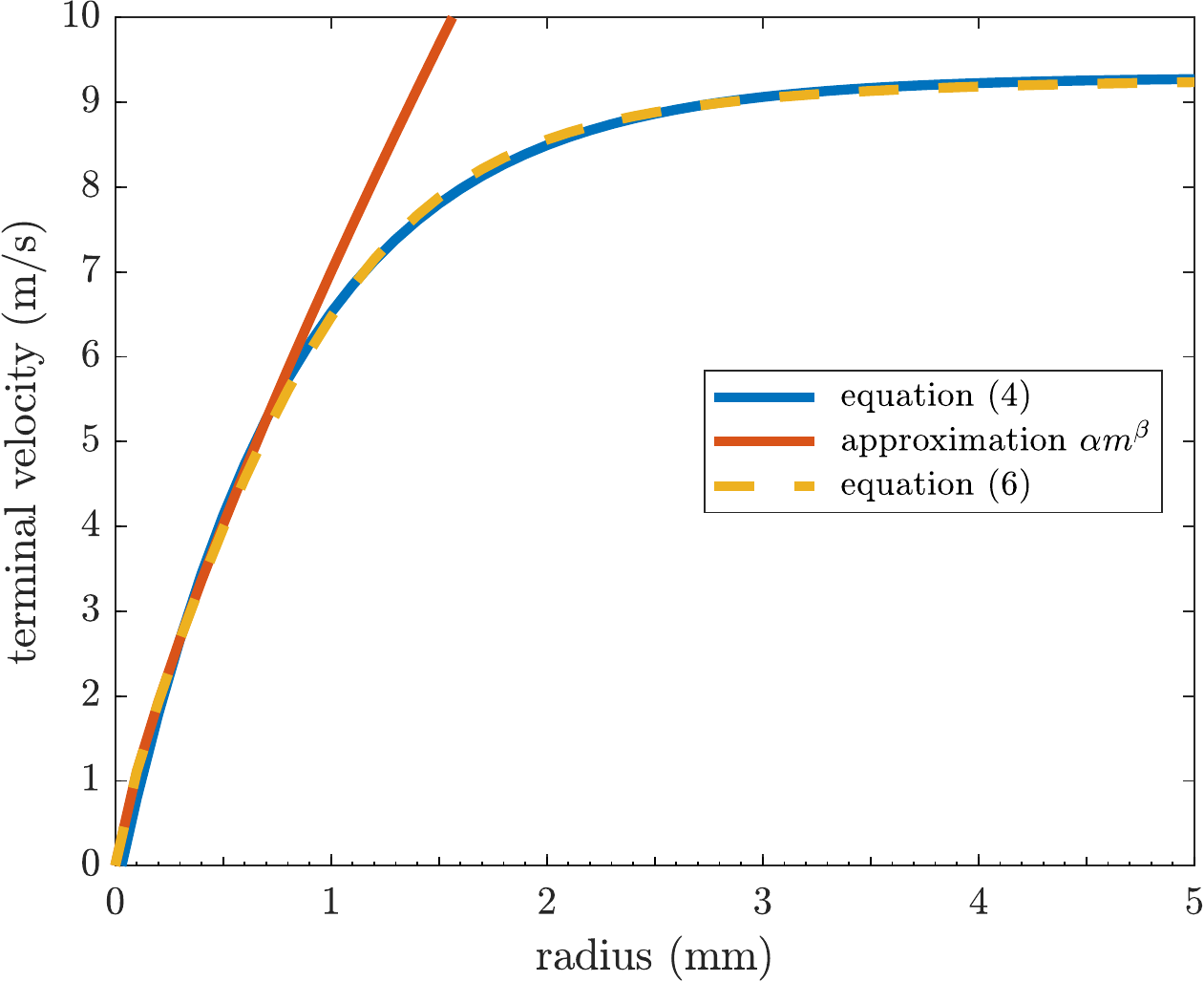}
  \caption{Terminal velocity $v_t$ as a function of the radius; values are calculated for the density $\rho=\rho_*$: Blue: ``Gold standard'' (\ref{eq:vt_ref}) from \citet{seifert_etal2014}; red: simple approximation $v_t=\alpha m^\beta$ in the spirit of \citet{seifert_beheng2006}; yellow: new approximation (\ref{eq:vt_final}).}
  \label{fig:vt}
\end{figure}

Here we propose the functional relation
 \begin{equation}
   \label{eq:vt_final}
   v_t(m)=\alpha\, m^\beta\, \left(\frac{m_t}{m+m_t}
   \right)^\beta
   \left(\frac{\rho_*}{\rho}\right)^{\frac{1}{2}}
 \end{equation}
 with $\alpha=\SI{190.3}{\meter\per\second\per\raiseto{\beta}\kilogram}$ and
 $m_t=\SI{1.21e-5}{\kilogram}$, which is similar to the
 \citet{seifert_beheng2006} model, but shows the same asymptotic behaviour as
 (\ref{eq:vt_ref}) for large rain drops, compare Figure~\ref{fig:vt}. 
 Using the approximation $m\approx q_r/n_r$ we thus obtain the formula
\begin{equation}
  \label{eq:vt_final_qrnr}
    v_t  =  \alpha\, q_r^\beta\, \left(\frac{m_t}{q_r+m_tn_r}
   \right)^\beta
   \left(\frac{\rho_*}{\rho}\right)^{\frac{1}{2}},
\end{equation}
to be used in our parameterization.

\subsubsection{Diffusional growth/evaporation for a single water
  particle}

The growth/evaporation of a single water particle of spherical shape
with radius $r$ and mass $m$ can be formulated as follows \citep[see,
e.g.,][]{pruppacher_klett2010}:
\begin{equation}
  \label{eq:growth_particle}
  \dv{m}{t}=-4\pi D r \rho(q_{vs}-q_v) G f_v\,,
\end{equation}
which involves the following terms:
\begin{itemize}
\item A diffusion constant \citep[cf.][]{pruppacher_klett2010}
  \begin{equation*}
    D=D_{0}\, \left(\frac{T}{T_0}\right)^{1.94}\frac{p_*}{p}
  \end{equation*}
  depending on temperature and pressure with
  \begin{equation*}
\begin{aligned}
  D_{0}=\SI{2.11e-5}{\square\meter\per\second},& &T_0=\SI{273.15}{\kelvin}.
\end{aligned}
\end{equation*}
\item The influence of latent heat release is given by
  \begin{equation*}
    G=\left[
      \left(\frac{L}{R_vT}-1\right)\frac{Lp_{s}}{R_vT^2}\frac{D}{K}
      + 1 
    \right]^{-1},
  \end{equation*}
  where the thermal conductivity $K$ is given\footnote{We have
    corrected a typo in the original formulation by \citet{dixon2007},
    after comparison with tabulated values of $K$, i.e., we have multiplied
    $a_K$ by a factor of $0.1$.} by \citep[][]{dixon2007}
    \begin{equation*}
      \label{eq:conductivity_air}
      K(T)=\frac{a_K\, T^\frac{3}{2}}{T+b_K\,
        10^{\frac{c_K}{T}}}
    \end{equation*}
    with
    % \todo{ich habe hier einfach die Konstante $a_k$ korrigiert,
    % d.h. $\cdot 0.1$ gegenüber Dixon (2007)}
  % \textbf{\color{red} ist $a_k$ richtig? Fehler in Dixon}
\begin{equation*}
\begin{aligned}
  a_K=\SI{0.002646}{\watt\per\meter\per\kelvin\raiseto{-\frac{3}{2}}\kelvin},&
  &b_K=\SI{245.4}{\kelvin},& &c_K=\SI{-12}{\kelvin}. 
\end{aligned}
\end{equation*}
\item A correction for ventilation effects: If a large particle, i.e.,
  a rain drop, is falling through air, vortices and turbulence are
  induced, which enhance evaporation
  \citep{pruppacher_rasmussen1979}. To account for this, an additional
  empirical ventilation coefficient $f_v$ is introduced in
  (\ref{eq:growth_particle}); according to \citet{seifert_beheng2006}
  we let
  \begin{equation*}
    \label{eq:fv}
    \begin{aligned}
    f_v=a_v+b_v\, N_{\mathrm{Sc}}^\frac{1}{3}N_{\mathrm{Re}}^\frac{1}{2},& &a_v=0.78,& &b_v=0.308 \,,
  \end{aligned}
\end{equation*}
  where the Schmidt and Reynolds numbers are defined as
  \begin{equation*}
    \label{eq:def_Nsc_Nre}
\begin{aligned}
    N_{\mathrm{Sc}}=\frac{\mu}{\rho D},& &
    N_{\mathrm{Re}}=\frac{2\rho r}{\mu}\,v_t 
    \,=\, \frac{2\rho}{\mu} \,v_t\,
    \left(\frac{3}{4\pi}\right)^\frac{1}{3}\rho_l^\frac{1}{3} m^\frac{1}{3},
  \end{aligned}
\end{equation*}
  in terms of the dynamic viscosity $\mu$ of air. The latter 
  can be expressed as a function of temperature \citep[cf.][]{dixon2007}, i.e.,
  \begin{equation*}
    \label{eq:mu}
\begin{aligned}
  \mu=\frac{\mu_0 T^\frac{3}{2}}{T+T_\mu},& &\mu_0=\SI{1.458e-6}{\second\pascal\per\raiseto{\frac{1}{2}}\kelvin},& &T_\mu=\SI{110.4}{\kelvin}.
\end{aligned}
\end{equation*}
\end{itemize}
For cloud droplets we can neglect ventilation effects, hence the mass rate
of a cloud droplet is given by
\begin{equation}
  \label{eq:growth_mc}
  \dv{m_c}{t}=-4\pi Dr \rho(q_{vs}-q_v) G =
%  \underbrace{4\pi \left(\frac{3}{4\pi \rho_l}\right)^\frac{1}{3}
%    DG}_{=d} \rho (q_v-q_{vs}) m_c^\frac{1}{3} =
  d\rho(q_v-q_{vs}) m_c^\frac{1}{3}
\end{equation}
with
\begin{equation}
\label{eq:d}
   d = 4\pi \left(\frac{3}{4\pi \rho_l}\right)^\frac{1}{3} DG.
\end{equation}
On the other hand, for rain drops we neglect condensation, hence
\begin{equation}
\label{eq:rainevaporation}
\dv{m_r}{t} =
%-4\pi \left(\frac{3}{4\pi \rho_l}\right)^\frac{1}{3}
%  DG
 % d \rho (q_{vs}-q_v)_+
 %  (a_E+b_E\,v_t(m_r)^\frac{1}{2}\,m_r^\frac{1}{6})m_r^\frac{1}{3}
   -d \rho (q_{vs}-q_v)_+
  \left[a_E\, m_r^\frac{1}{3}+ b_E\,v_t(m_r)^\frac{1}{2}
    m_r^\frac{1}{2} \right] ,
\end{equation}
where
\begin{equation}
  \label{eq:aEbE}
  \begin{aligned}
  a_E=a_v=0.78,& &b_E=b_v\left( \frac{\mu}{\rho D}\right)^\frac{1}{3}
    \sqrt{\frac{2\rho}{\mu}}
    \left(\frac{3}{4\pi \rho_l}\right)^\frac{1}{6}.
  \end{aligned}
\end{equation}
Here we have used the short-hand notation
\begin{equation*}
  (q_{vs}-q_v)_+ = \begin{cases} 
    q_{vs}-q_v, & q_v\leq q_{vs} \ \text{(subsaturated regime)},\\
    0, & q_v > q_{vs} \ \text{(supersaturated regime)}.
  \end{cases}
\end{equation*}

\begin{remark}
  A more general approach suggested, e.g., in \citep{pruppacher_klett2010}
  also includes a kinetic correction of the diffusivity $D$ 
%and the  conductivity $K$ 
  for very small droplets. Since
  calculations show that these corrections are not relevant for cloud
  droplets with a radius $r>\SI{5}{\micro\meter}$ we
  omit these corrections in this study. % Such a correction also does not
  % solve the lack of Lipschitz continuity in the limit
  % $m\rightarrow 0$ because the right-hand side of (\ref{eq:singular})
  % still exhibits powers $m^a$ with exponents $a<1$.
\end{remark}

\subsubsection{Collision of rain drops with cloud droplets and
   coalescence/accretion}
\label{sec:single_rate_accretion}

A spherical rain drop of radius $r$ and mass $m_r$ falls with terminal
velocity $v_t(m_r)$ through a cylindrical volume 
\begin{equation*}
  V=\pi r^2 \, v_t(m_r)\Delta t=
  \pi \left(\frac{3}{4\pi \rho_l}\right)^{\frac{2}{3}} m_r^\frac{2}{3}
  v_t(m_r) \Delta t 
\end{equation*}
during a time interval $\Delta t>0$. Within this volume there is a total mass 
$M_c=V\rho q_c$ of cloud droplets that will be hit by this rain drop.
The corresponding mass growth rate of the rain drop is thus given by
\begin{equation}
\label{eq:k3prime}
  \dv{m_r}{t} = k_2' V \rho q_c
                    = k_2' \rho q_c \pi
                    \left(\frac{3}{4\pi \rho_l}\right)^\frac{2}{3}
                    m_r^\frac{2}{3}
                    v_t(m_r),
\end{equation}
where $k_2'>0$ is the associated efficiency parameter.

%{\blue 
\subsection{Computing mass/number concentration rates from
single particles rates}

In this section we derive rates of change of the bulk
quantities. For this purpose rates for single particles are
scaled up with the corresponding particle  number concentration. % For cloud droplets,
% on the other hand, we do not keep track of the number concentration $n_c$;
% therefore we have to derive the mass concentration rate of cloud droplets
% somewhat differently.

\subsubsection{Rates for diffusional growth/evaporation of rain drops}

Evaporation of rain drops affects mass concentration but also number
concentration, because small droplets may evaporate completely. We 
therefore use
\begin{equation*}
  E = n_r\, \eval{\dv{m_r}{t}}_{\text{evaporation}}
\end{equation*}
for the evaporation term of the mass concentration rate, and we assume that
the reduction of the number concentration is proportional to the evaporated
mass; the proportionality factor is set to be equal to the inverse of 
the average mass $\overline{m}_r$ of rain drops. Accordingly, we let
\begin{equation}
\label{eq:Eprime}
  E' = \frac{1}{\overline{m}_r}E
\end{equation}
be the corresponding number concentration rate.
%, where $\overline{m}_r=q_r/n_r$ 
%is the average mass of all existing rain drops.
Inserting (\ref{eq:rainevaporation}) we thus obtain
\begin{equation}
\label{eq:E}
  E %= n_r\, \dv{m_r}{t}\Big|_{\text{evaporation}} 
    = -d\rho (q_{vs}-q_v)_+
                             \left( a_E q_r^{\frac{1}{3}}n_r^{\frac{2}{3}}
                             + b_Ev_t^{\frac{1}{2}}q_r^{\frac{1}{2}}n_r^{\frac{1}{2}}
                                 \right)
\end{equation}
and 
\begin{equation*}
  E' %= \frac{1}{\overline{m}_r}\, E 
     = -\frac{1}{\overline{m}_r}d\rho (q_{vs}-q_v)_+
                                  \left( a_E q_r^{\frac{1}{3}}n_r^{\frac{2}{3}}
                                  + b_Ev_t^{\frac{1}{2}}q_r^{\frac{1}{2}}n_r^{\frac{1}{2}}
                                 \right),
\end{equation*}
where $v_t$ is given by \eqref{eq:vt_final_qrnr}.

\subsubsection{Rates for the accretion of rain drops}

Concerning the collision processes between rain drops and cloud
droplets we obtain in the same way
%start by considering a single rain drop during sedimentation as 
%in section~\ref{sec:single_rate_accretion}. Integrating over all available 
%rain drops we obtain 
the corresponding mass concentration rate
\begin{equation}
  \label{eq:A2}
  \begin{aligned}
  A_2 &= n_r \eval{\dv{m_r}{t}}_{\text{accretion}}
  = n_r k_2 \rho q_c \pi
  \left(\frac{3}{4\pi \rho_l}\right)^{\frac{2}{3}}
  m_r^{\frac{2}{3}} v_t\\
  &= k_2 \pi
  \left(\frac{3}{4\pi \rho_l}\right)^{\frac{2}{3}}
  v_t\rho q_cq_r^{\frac{2}{3}}n_r^{\frac{1}{3}},
\end{aligned}
\end{equation}
where we have approximated $m_r\approx q_r/n_r$ in the last step.
Note that this is a simplistic result which does not take into account,
for example, that the average velocity of rain drops is not equal to the
terminal velocity of a rain drop with average mass; but this can be
compensated by calibrating the efficiency parameter $k_2$, which may be
somewhat different from $k_2'$ in (\ref{eq:k3prime}). 
%accomodated for by that 
%with the terminal velocity $v_t$ and a potential correction
%factor $c_q$, which takes into account that available measurement data of rain
%droplet sizes exhibit skew distributions with long tails;
%  are similar to an exponential distribution, which is skew to larger sizes; 
%thus, the factor $c_q>1$ corrects for the higher
%weight of droplets which are larger than the mean value (see also
%section~\ref{sec:sedimentation}).
\begin{remark}
  For smaller rain drops one can approximate 
  $v_t\approx \alpha(q_r/n_r)^\beta
   =\alpha q_r^\beta n_r^{-\beta}$ with $\beta=4/15$, cf.~(\ref{eq:vt_sb06}), 
  and then
  \begin{equation*}
    A_2\sim q_c q_r^{\frac{2}{3}+\beta}n_r^{\frac{1}{3}-\beta}=
    q_c q_r^\frac{14}{15}n_r^\frac{1}{15}.
  \end{equation*}
  This is almost equivalent to a standard predator-prey
  formulation $A_2\sim q_cq_r$
  with rain drops as predator population, depleting the
  prey population of cloud droplets \citep[see, e.g.,][]{wacker1992}.
%% MHB: deleted following sentence (gilt fuer alle unsere Parameter)
%  The corresponding proportionality constant could be estimated from
%  measurements or from benchmark simulations using more sophisticated 
%  cloud models.
\end{remark}

\subsubsection{Sedimentation of rain drops}
\label{sec:sedimentation}
So far all the considered processes take place within each individual 
control volume. Sedimentation, on the other hand, produces a flux through
the boundaries of the control volumes.
Let $z$ be the coordinate of the vertical position (above sea level) of
the control volume.
As stated above rain drops accelerate due
to gravity to a reasonable terminal velocity.
This will be used to derive fluxes for the bulk variables mass and number
concentrations of the rain drops to specify their vertical advection. 
Distinguishing between effective velocities 
$v_{q}$ and $v_{n}$ for mass and number concentrations, respectively,
the corresponding fluxes $J$ and $J'$
% for mass and number concentration, respectively,
are given by
\begin{equation*}
\begin{aligned}
  J=v_{q}\,\rho q_r,& &J'=v_{n}\, \rho n_r
\end{aligned}
\end{equation*}
with units
\begin{equation*}
\begin{aligned}
  [J]=\si{\kilogram\per\square\meter\per\second},& &[J']=\si{\per\square\meter\per\second}.
\end{aligned}
\end{equation*}
The effective velocities $v_{q}$ and $v_{n}$ correlate with the terminal
velocity $v_t$ of a single drop with average mass, i.e., we let
\begin{equation}
\label{eq:vqvn}
\begin{aligned}
v_{q}=c_qv_t,& &v_{n}=c_nv_t, 
\end{aligned}
\end{equation}
with parameters
\begin{equation}
\label{eq:cqcn}
%  c_q>1>c_n>0.
  c_q>c_n>0.
\end{equation} 
The weight $c_q$ takes into account that
the size distribution of rain drops is often observed to have heavy tails
\citep{marshall_palmer1948},
and that larger drops contribute more to the mass sedimentation flux 
than smaller ones. In contrast, drops smaller than the mean size
yield the dominant contribution to the sedimentation number flux. 
%therefore we assume that $c_n<1$. 
In short, one can say that more larger drops than smaller ones
fall out of a box, and this is taken into account by our 
constraints~(\ref{eq:cqcn}) on the parameters $c_q$ and $c_n$ 
\citep[see, e.g., the discussion
in][]{wacker_seifert2001}. 
%% Referee: wants more evidence for eq.~{eq:cqcn}

\begin{remark}
  The condition~(\ref{eq:cqcn}) is fulfilled in a natural way when using the
  exponential or the Gamma distribution for the mass distribution of
  rain drops \citep[as suggested, e.g., in][]{seifert_beheng2006},
  because this is equivalent to the inequality of moments
  $\mu_{\beta+1}\mu_0\ge \mu_\beta\mu_1$ of the distributions with
  $\beta\in\mathbb{R}_+$; this is true, since $\log\mu_r$ fullfills
  Lyapunov's inequality \citep{book_marshall_etal2011}.
\end{remark}

In the column model the sedimentation terms 
\begin{equation*}
\begin{aligned}
  S=\frac{1}{\rho}\pdv{J}{z},& &S'=\frac{1}{\rho}\pdv{J'}{z}
\end{aligned}
\end{equation*}
appear as sources and sinks, respectively, in the time evolution,
and turn the overall model into a  
hyperbolic system of partial differential equations.
For the box model without flux from above, we can simplify
the sedimentation terms $S$ and $S'$ by using the vertical extension $h$ of
the box to obtain
\begin{equation}
\label{eq:Sout}
\begin{aligned}
  S = S_{\text{out}}=\frac{J}{h\rho}=\frac{v_qq_r}{h}\,,& &S'= S_{\text{out}}'=\frac{J'}{h\rho}=\frac{v_nn_r}{h}\,.
\end{aligned}
\end{equation}
Note, that the height of the box may change with time due to
adiabatic expansion or compression. See also the discussion in
Section~\ref{subsec:solvability}.

%\textbf{still need to introduce z somewhere!}

\subsection{Collision of cloud droplets -- autoconversion}
In a control volume $\Delta V$ the volume fraction
occupied by cloud droplets is given by
%$\rho q_c\Delta V$, representing a volume of liquid water of
$\rho q_c/\rho_l$.
The probability that any single cloud droplet collides with any other
cloud droplet and that they recombine to a rain drop 
-- called autoconversion --
is proportional to the size of this volume fraction.
It follows that 
\begin{equation}
\label{eq:autorate1}
  k_1\frac{\rho q_c}{\rho_l}\,\Delta t
%,~~k_1'\in\mathbb{R}_+,~~[k_1']=\SI{}{s^{-1}}
\end{equation}
is the expected number of autoconversions of an individual cloud
droplet in a sufficiently small time interval $\Delta t$, where $k_1$
with $[k_1]=\si{\per\second}$ is the corresponding proportionality
constant. Note that we are only interested in those collisions, which
result in a single drop which is large enough to be registered as a
new rain drop, because the other collisions have no effect on our
concentration variables. In addition, the effect of such collisions is
quite small.

Multiplying eq.~(\ref{eq:autorate1})
with the total number of cloud droplets $\rho n_c\Delta V$ in the
control volume we get the number of autoconversions
\begin{equation*}
  \frac{1}{2}k_1(\rho n_c)\frac{\rho q_c}{\rho_l}\Delta t\Delta V
\end{equation*}
in $\Delta V$ within the time interval $\Delta t$;
the factor $\frac{1}{2}$ prevents double counting of events. 

Since each autoconversion recombines two cloud droplets into a new rain drop,
the corresponding rate of the number concentration of rain drops 
per mass of dry air (i.e., $\rho\Delta V$) is given by
% . Using
%the mass of the air parcel (mass of dry air) as new reference, we
%obtain the rate of number concentration
\begin{equation*}
  A_1'=k_1\frac{\rho n_cq_c}{2\rho_l}.
\end{equation*}
On the other hand, each collision increments the mass of rain drops by
two times the average mass $\overline{m}_c=q_c/n_c$ of cloud droplets,
% forms a new rain drom two droplets with average mass 
%$\overline{m}_c=q_c/n_c$ are removed from the cloud droplets
%population, 
which leads to the autoconversion rate for the rain drop mass concentration
\begin{equation}
\label{eq:A1}
  A_1=2k_1\overline{m}_c\frac{\rho n_cq_c}{2\rho_l}
     =k_1\frac{\rho q_c^2}{\rho_l}.
%\sim q_c^2
\end{equation}

% %PS hier muss man entscheiden, ob man das in den Text reinbringt
% \noindent\textbf{Remark:}\\
% In our formulation we omit so-called selfcollection of cloud droplets;
% this means that cloud droplets collide but the resulting droplet is
% not large enough to be counted as rain drop. This process, as
% described, e.g., by \citet{seifert_beheng2006} does not affect $q_c$
% but would change droplet number $n_c$; however, this would not be
% consistent with our formulation of condensation. In addition, the
% effect of this process is quite small, therefore we decided to omit
% this process.
% %PS

\subsection{Treatment of cloud droplet condensation}
\label{sec:cloud_condensation}
The treatment of cloud droplet condensation leads to several subtle
issues, which must be considered carefully in the development of a
consistent and numerically tractable scheme.

\subsubsection{Particle formation}
\label{subsubsec:particle_formation}
In the atmosphere many aerosol particles 
are available. Some of them,
depending on their chemical components (i.e., their hygroscopicity), 
have the ability to attract water molecules.
As soon as there is water vapour these particles grow by diffusion, i.e., 
a phase transition from the gas phase to mixed particles including liquid water
takes place. This effect can be
described by K\"ohler theory \citep[see, e.g.,][]{koehler1936,
  pruppacher_klett2010}, which determines the size of the grown aerosol at
a given saturation ratio $q_v/q_{vs}$ in dependence of the initial
size of the aerosol and its chemical properties. A more compact
formulation of this theory can be found in
\citet{petters_kreidenweis2007}, using the hygroscopicity as single parameter. 
The K\"ohler theory predicts a so-called critical radius $r_0$ and
a maximal supersaturation ratio $S_0=q_v/q_{vs}>1$, such that there is a
one-to-one relation between $q_v<S_0q_{vs}$ and the radius $0<r<r_0$
of a given wetted aerosol.
Once the saturation has reached the critical level $S_0$
the aerosol particle becomes unstable, i.e.,
it can grow to (almost) arbitrarily large sizes; 
%Therefore, once the critical radius $r_0$ has been reached, 
this grown aerosol particle can now be called a cloud droplet.  

There are complex cloud models, which try to
take into account the complicated procedure of activation, but most
standard cloud models do not consider aerosol particles, so that the
generation or activation of droplets must be treated in a somewhat
artifical manner.
Often, for example, a certain number of cloud droplets is activated, 
once a certain threshold of supersaturation is reached.
In even simpler models (mostly single moment schemes), 
it is often assumed that in case of
supersaturation all aerosol particles are activated instantaneously,
i.e., $n_c$ is set to the number concentration of available CCN
\citep[see, e.g.,][]{grabowski1999}; likewise the number concentration
is set to zero in subsaturated conditions.

\begin{figure}[t!]
  \centering
  \includegraphics[height=0.505\linewidth]{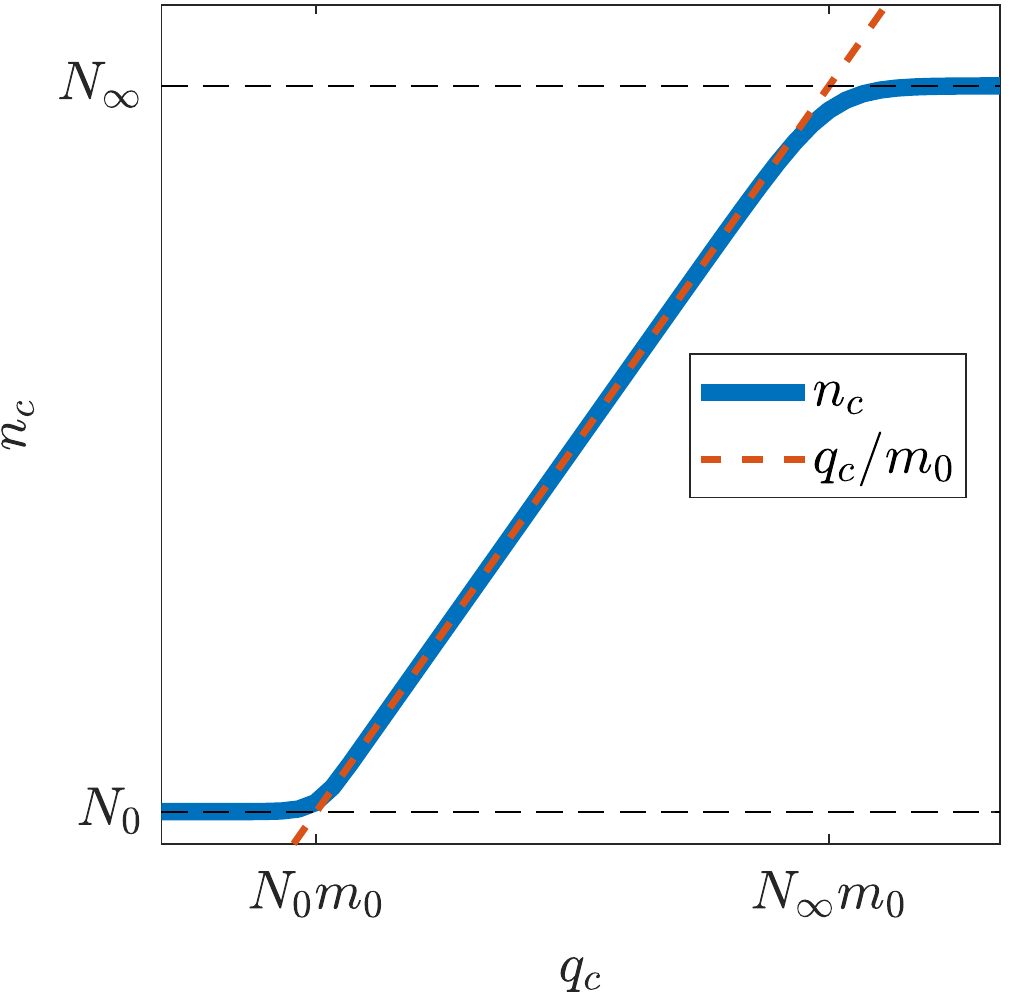}
  \caption{Number concentration $n_c$ depending on the mass
    concentration $q_c$ as given in equation~(\ref{eq:nc_relation}).} 
  \label{fig:nc_qc}
\end{figure}

In our model we assume that there are $N_0$ aerosols (per volume and mass of
dry air) which reach the
critical K\"ohler radius first, e.g., because they are largest.
This implies that the counting concentration $n_c$ is 
already positive before the first
%at least $N_0$ as soon as 
droplets can be registered, i.e., when $q_c$ becomes positive.
Later, other aerosols may turn into droplets, and we further assume,
that there are at most $N_\infty$ aerosols (per volume and mass of dry air)
available.
To be specific, we assume that $n_c$ is a function of $q_c$, namely
\begin{equation}
  \label{eq:nc_relation}
  n_c=\frac{q_cN_\infty}{q_c+N_\infty m_0}
  \coth(\frac{q_c}{N_0m_0})  
\end{equation}
with free parameters $N_\infty,N_0,m_0$, see Figure~\ref{fig:nc_qc}.

This function represents three different regimes: ($i$) Before the maximal
saturation level $S_0$ is reached, small aerosol particles are already around, 
but the total liquid droplet mass concentration is still negligible 
(i.e., $q_c=0$). However, the number
concentration is already equal to the parameter $N_0$.
($ii$) At growing supersaturation, more and more cloud droplets appear,
and all aerosols compete for the available
water vapour, so that the mean size of all particles is approximately
constant. Therefore, in this regime there is an approximately linear relation 
between $n_c$ and $q_c$, i.e., $q_c\approx m_0 n_c$. In particular, 
this implies that we have an ongoing in-cloud activation of new cloud droplets
with increasing saturation rate. The parameter
$m_0$ can be interpreted as the typical water mass content
of a cloud droplet
close to activation. 
($iii$) At high supersaturation levels all CCN are activated, thus the 
droplet number concentration is almost equal to the total number of CCNs, 
i.e., $n_c\approx N_{\infty}$. 

We will demonstrate in Section~\ref{subsec:Newton} below
that this nonlinear coupling of the droplet number and mass concentrations
%model, which is based on physical understanding,
entails an automatic (i.e., implicit) particle activation.
%, as we will demonstrate later in Section~\ref{subsec:Newton}.
By changing the tuning parameters $N_0$, $N_\infty$, and $m_0$ it is possible
to represent different aerosol regimes (i.e. polluted, clean,
maritime regimes; compare Section~\ref{subsec:activation}).
%Moreover, our approach allows for supersaturated regimes.

\begin{remark}
  We found that the algebraic constraint~(\ref{eq:nc_relation}) is
  better suited for modelling the activation of cloud particles on
  physical grounds than any of the differential equations for $n_c$ as
  a function of time that we could think of.
%  With our activation parameterisation, the rate equation for the
%  number concentration can be obtained from the equation for the mass
%  concentration by scaling with mean mass $m_c=q_c/n_c$. For clouds we
%  can assume in a good approximation that the mean mass is
%  approximately constant. We find the same scaling for collision
%  processes treated in the formulations below. Thus, we can reduce the
%  system of equations by one equation, i.e. we have only differential
%  equations for cloud variables $q_c,q_r,n_r$ together with the
%  algebraic equation for $n_c$
\end{remark}

\subsubsection{Condensation rate for cloud droplets}
\label{subsubsec:explicit_growth}

%In the temperature range $\SI{250}{K}\le T\le \SI{310}{K}$ 
In warm clouds the amount
of available water molecules in the gas phase is very high and the
diffusivity is quite large, hence diffusional growth of droplets is a
very fast process, if cloud droplets are already available. Therefore
supersaturation due to cooling of air, for example, changes very
rapidly towards thermodynamic equilibrium, i.e., $q_v\approx q_{vs}$.
%  This feature can also be measured inside clouds. 
Accordingly, many cloud models use saturation adjustment, 
which means that for $q_v>q_{vs}$ all
%assume for the water
%  vapour concentration inside clouds $q_v=q_{vs}$ and use the
%  so-called technique of saturation adjustment; for $q_v>q_{vs}$ all
excess water vapour is instantaneously turned into cloud droplet mass
concentration $q_c$, so that $q_v=q_{vs}$.
%; the released latent heat is also instantaneously used
% for changing the temperature.

Saturation adjustment can be solved numerically in a very efficient
way by using Newton's method \citep[][]{kogan_martin1994}. However,
the method leads to some problematic phenomena. First, as we have discussed
in Section~\ref{subsubsec:particle_formation}, the activation of
droplets inside clouds is nonphysical, strictly speaking, since this
activation requires supersaturation.  Therefore activation of cloud
droplets has to be carried out separately before saturation adjustment
is performed.
% Accordingly, the cloud processes have to be ordered to
% accomodate for the needs of the numerical algorithms rather than their
% physical justifications, which can result in inconsistencies.
Second,
saturation adjustment has been shown to lead to an overestimation of
latent heat release during condensation, because all excess water
vapour is turned into liquid water at once. This yields higher
buoyancy and introduces errors in the representation of systems with
convective updrafts \citep[see,
e.g.,][]{grabowski2015,grabowski_morrison2016,grabowski_morrison2017}.
Therefore saturation adjustment should be avoided whenever possible,
and explicit supersaturation regimes should be tolerated in modern
cloud models; see also Section \ref{subsec:versus_sat_adjust}.

As we have explained before, our model does allow supersaturation, and the
growth rate of the mass concentration $q_c$ follows from the growth equation
(\ref{eq:growth_mc}) to be
\begin{equation}
\label{eq:C}
   C = n_c\dv{m_c}{t} = d\rho(q_v-q_{vs})n_c^{\frac{2}{3}}q_c^{\frac{1}{3}},
\end{equation}
where we have taken $m_c=q_c/n_c$ to be the average mass of a cloud droplet.

\subsection{Full model equations: Box model}
\label{subsec:boxmodel}
The cloud model variables have to be coupled to thermodynamics,
i.e., to changes in pressure and temperature, respectively.
For this we assume adiabatic changes (no heat exchange
with the environment) when the control volume is moving vertically. The
adiabatic lapse rate $\gamma=g/c_p$ is used for these temperature changes. The
latent heat of a phase transition (water vapour $\leftrightarrow$ liquid
water) is also distributed in the volume, changing temperature. In
addition, we assume hydrostatic pressure change ($\pdv{p}{z}=-g\rho$),
which is a common assumption \citep[e.g.,][]{korolev_mazin2003}.

Now we can formulate the system of equations for a box model approach, given
the different sinks and sources of the water quantitities as described above.
We also assume some external forcing in terms of a (given)
vertical upward motion with velocity $w=w(t)$;
the latter can be used to model, e.g., the passage
over a mountain ridge or the ascent of a warm front onto cold air;
see Section~\ref{subsec:showcasing} for examples.
The system is given by
\begin{equation}
\label{eq:ode}
	\begin{array}{llrrrll}
	  \dot{q}_v & = & -C &      &      & +E, &\\
	  \dot{q}_c & = &  C & -A_1 & -A_2, &    &\\
	  \dot{q}_r & = &    &  A_1 & +A_2\phantom{,} & -E & -S\,,\\
	  \dot{n}_r & = &    &  A'_1&      & -E'& -S',\\[0.1cm]
	  \dot{p}   & = & \multicolumn{5}{l}{-\gamma \rho w,} \\
	  \dot{T}   & = & \multicolumn{5}{l}{-\gamma w - 
                          {\displaystyle\frac{L}{c_p}}(E-C),}
	\end{array}
\end{equation}
compare Figure~\ref{fig:processes}.
Its right-hand side depends on intermediate quantities,
but also on the coupled
number concentration $n_c$ of cloud droplets and on the density $\rho$ of
dry air; therefore, the system is
closed using the corresponding algebraic constraints
% % Porz : korrigierte Formel
\begin{eqnarray}
  \label{eq:nc}
  n_c & = & \frac{q_cN_\infty}{q_c+N_\infty m_0}
            \coth(\frac{q_c}{N_0 m_0}), \\
  \label{eq:rho}
  \rho & = & \frac{p}{R_aT}.
\end{eqnarray}

\subsection{Solvability of the differential equations}
\label{subsec:solvability}
Since the right-hand side of the differential equation~(\ref{eq:ode}) 
is not differentiable, no higher order regularity of the solution can
be expected. Moreover, the Picard-Lindel\"of theory is not applicable,
so that the differential equation has no unique solution, in general. 
The existence of solutions is nevertheless guaranteed by 
Peano's theory~\citep{walter1998}. 

The lack of uniqueness is apparent from
the differential equation for $q_c$: If the system is in the subsaturated 
regime, i.e., if $q_c$, $n_r$, and $q_r$ are zero at time $t=0$, then all
the driving terms $A_1$, $A_1'$, $A_2$, $C$, $E$, $E'$, $S$, and $S'$ on the 
right-hand side of (\ref{eq:ode}) vanish, and hence, the constant functions
$q_v=q_v(0)$, $q_c=0$, $q_r=0$, and $n_r=0$, solve the first four
differential equations in (\ref{eq:ode}) -- even if the system changes to
the supersaturated regime at some later time $t=t_0$ 
for a specific choice of upward drift $w$.
However, as will be shown in Section~\ref{subsec:Newton},
as soon as $q_{vs}<q_v$ our specific ansatz for generating cloud
droplets allows for a nontrivial solution of the system~(\ref{eq:ode}).

In our box model we treat the total mass $m_a$ of dry air within the box
as being constant over time, and we also freeze the horizontal cross
section $A$ of the box.  According to the gas law~(\ref{eq:gaslaw}),
however, the density $\rho$ may vary with time. We therefore need to
adapt the vertical extent $h=h(t)$ of the box to account
for changes in the density and to conserve the total air mass
$m_a=\rho A h$ over time.

\subsection{Formulation of a mass conserving column model}
\label{subsec:column_model}
The model of Section~\ref{subsec:boxmodel} 
can readily be extended to a vertical column of air in order to 
treat nontrivial vertical humidity distributions. 
To this end multiple boxes 
are stacked on top of each other. Strictly speaking, the column model consists
of an initial value problem for a hyperbolic differential algebraic system,
where we consider a Lagrangian air column with internal
sedimentation flow. The individual boxes provide a natural finite volume 
discretization in space; the time discretization will be worked out in more 
detail in Section~\ref{sec:numericalscheme}.

Concerning the conservation of mass (of dry air and of water, respectively)
we assume that the horizontal cross section of all boxes is the same and that
its area is $A$, and as for the box model we adapt the height $h=h(t,z)$ 
of each individual box over time to conserve the mass $m_a=m_a(z)$ of dry
air within every individual box;
the mass may, however, depend on the spatial variable $z$, i.e., be different
for each box. 
As we will see in Section~\ref{subsec:mass} this way we not
only conserve the mass of air, but also the total mass of water within
the column -- except for precipitation, of course.

\section{Numerical time integration}
\label{sec:numericalscheme}
%% MHB: deleted explicit mentioning of p and T below (no reason to do otherwise)
Starting from 
%given vertical profiles $p(0,z)$, and $T(0,z)$
%for pressure and temperature, respectively, and 
(consistent) initial values for the variables of our model at time $t=0$, 
the overall column model system with a given forcing velocity profile $w=w(t)$ 
can be integrated by stepping forward explicitly in time. 
%Starting from given vertical mass and mass profiles $m_a(z)$ and 
%$\rho(0,z)$ of dry air, (consistent) initial values for all the other 
%variables at time $t=0$, and a forcing velocity profile $w=w(t)$,
%the overall system can be integrated by stepping forward explicitly in time. 
For the hyperbolic column model this calls for a Courant-Friedrichs-Lewy (CFL)
condition, 
i.e., an upper bound of the time step $\tau>0$. In our context this constraint 
has an obvious physical interpretation: The inflow of falling rain drops 
(both, in terms of mass and number) into any given box must not traverse 
this box within a single time step, so that the flows
across all horizontal box faces are independent of each other for every fixed
time step. Because of our assumption $v_q>v_n$, 
see (\ref{eq:vqvn}), (\ref{eq:cqcn}), this amounts to the upper bound
\begin{equation*}
\label{eq:CFL}
\begin{aligned}
  \tau \,<\, \min\{h/v_q\},& &v_q = c_q v_t,
\end{aligned}
\end{equation*}
the minimum being taken over all boxes at a given time step.

In order to maintain nonnegativity of all water concentrations, 
see Sections~\ref{subsec:qr12nr12} and \ref{subsec:Newton} for
more details, we split the sedimentation term into its outflow
and inflow components,  $S_{\text{out}}$ and $S_{\text{in}}$, respectively, 
and treat them separately as sinks and sources.
In this manner the resulting overall system can be considered an ordinary 
differential algebraic system.
For the numerical treatment it is important that this differential algebraic
system has index one, i.e., that the closing conditions (\ref{eq:rho}) 
and (\ref{eq:nc}) can be solved (explicitly) for the 
algebraic variables $\rho$ and $n_c$.
By updating these algebraic variables in each time step after all other 
variables -- except for $h$, see Section~\ref{subsec:mass} below -- 
we make sure that the two algebraic constraints are consistent 
after each individual time step.

Concerning the other variables we use a semi-implicit Euler scheme, 
as worked out in detail in the following sections, where
certain variables are treated explicitly, while other variables
are solved for implicitly, but very efficiently.
At the beginning of each time step we evaluate all the 
parameters of the different processes,
i.e., the saturation vapour concentration (\ref{eq:vs}), 
the terminal velocity (\ref{eq:vt_final}), the diffusivity (\ref{eq:d}),
and the ventilation parameter $b_E$ of (\ref{eq:aEbE}), using
the values of the depending variables from the previous time step. 

\subsection{The semi-implicit Euler scheme}
\label{subsec:Euler}
We split the right-hand side of the differential equation (\ref{eq:ode}) 
into two parts, one of which is being treated implicitly, while the other one
is treated explicitly. For both parts we use the Euler scheme, 
because the solution lacks regularity in general; 
see Section~\ref{subsec:solvability}. The implicit part contains
the entire right-hand side for the cloud droplet mass concentration and includes all 
the sinks for the rain drop mass and number concentrations. 
This splitting ensures:
\begin{itemize}
\item[(a)] an adequate activation of cloud droplets as soon as the system
           becomes supersaturated, and
\item[(b)] that all water concentrations remain nonnegative.
\end{itemize}

To be specific, starting from the current values $q_{v,i}$, $q_{c,i}$,
$q_{r,i}$, $n_{r,i}$, $n_{c,i}$, $p_i$, $T_i$, and $\rho_i$ of the approximate
solution of \eqref{eq:ode}-\eqref{eq:rho} in some given box
at time $t_i=i\tau$, we first solve
\begin{align}
\label{eq:qr12}
   q_{r,i+1/2} =&\ q_{r,i} - 
     \tau\qty(E(q_{r,i+1/2},n_{r,i+1/2}) \,+\,
        S_{\text{out}}(q_{r,i+1/2})),\\
\label{eq:nr12}
   n_{r,i+1/2} =&\ n_{r,i} -  
     \tau\qty(E'(q_{r,i+1/2},n_{r,i+1/2}) \,+\,
        S'_{\text{out}}(n_{r,i+1/2})),
\end{align}
for $q_{r,i+1/2}$ and $n_{r,i+1/2}$. Then we determine $q_{c,i+1}$ from
\begin{equation}
\label{eq:qci1}
   q_{c,i+1} = q_{c,i} + \tau\qty(C(q_{c,i+1}) \,-\, A_1(q_{c,i+1}) \,-\, 
              A_2(q_{c,i+1},q_{r,i+1/2},n_{r,i+1/2})),
\end{equation}
and finally, we update the new values of $q_r$ and $n_r$ as
\begin{align}
\label{eq:qri1}
   q_{r,i+1} =&\ q_{r,i+1/2} + \tau \qty(A_1(q_{c,i+1}) \,+\, 
              A_2(q_{c,i+1},q_{r,i+1/2},n_{r,i+1/2}) \,+\, S_{\text{in}}),\\
\label{eq:nri1}
   n_{r,i+1} =&\ n_{r,i+1/2} + 
             \tau \qty(A_1'(q_{c,i+1}) \,+\, S'_{\text{in}}).
\end{align}
%% MHB: interchanged the next two sentences
We use the old values $q_{v,i}$, $n_{c,i}$, $\rho_i$, $p_i$, and $T_i$, 
when evaluating the respective terms on the right-hand sides of 
(\ref{eq:qr12})-(\ref{eq:nri1}).
In (\ref{eq:qri1}) and (\ref{eq:nri1}) the inflows $S_{\text{in}}$ and 
$S'_{\text{in}}$ are given by the corresponding outflows of the neighboring box, 
which have been determined in steps (\ref{eq:qr12}) and (\ref{eq:nr12}).
This implies that in the implementation of the column model each update
\eqref{eq:qr12}-\eqref{eq:nri1} should be done simultaneously for all 
individual boxes to have the inflows available when needed; note that this
allows for a straightforward SIMD parallelization 
(single instruction, multiple data) of the column model.

In Section~\ref{subsec:qr12nr12} we show that the system (\ref{eq:qr12}), 
(\ref{eq:nr12}) has a unique nonnegative solution $q_{r,i+1/2}$, $n_{r,i+1/2}$,
which can be written down explicitly.
Step (\ref{eq:qci1}), on the other hand, is treated in 
Section~\ref{subsec:Newton}: 
%The solution of (\ref{eq:qci1}) 
It can be reduced 
to the computation of a specific root of a polynomial of degree six, 
and a straightforward implementation of the Newton method provides a very
efficient scheme for the computation of $q_{c,i+1}$, with guaranteed quadratic
convergence.

\begin{remark} 
In the important special case, where a cloud parcel becomes 
supersaturated, but no cloud droplets do yet exist,
the corresponding solution $q_{c,i+1}$ will be positive; compare item (a)
in Section~\ref{subsec:Newton}.
Therefore the implicit Euler step (\ref{eq:qci1}) is our key means to 
automatically invoke the activation of cloud droplets discussed 
in Section~\ref{subsubsec:particle_formation}.
\end{remark}

Once the new values for $q_c$, $q_r$, and $n_r$ have been computed, 
the pressure $p$, the temperature $T$, and the vapour concentration $q_v$ 
are updated with an explicit Euler step, using the quantities
\begin{equation*}
  \begin{aligned}
   E = E(q_{r,i+1/2},n_{r,i+1/2})& &\text{and}& &
   C = C(q_{c,i+1})
 \end{aligned}
\end{equation*}
determined in (\ref{eq:qr12}) and (\ref{eq:qci1}), respectively.
Finally the algebraic variables $\rho$ and $n_c$ are updated, 
and the new box heights are retrieved from the identity
\begin{equation}
\label{eq:hi1}
	h_{i+1} = \frac{m_a}{\rho_{i+1}A}\,,
\end{equation}
where the mass $m_a$ and the area $A$ of the horizontal cross section of
each box stay constant over all times. This ensures conservation of the
mass of dry air.

\subsection{Water mass conservation}
\label{subsec:mass}
Assuming that there is no inflow into the column (box) from above,
it is obvious from (\ref{eq:ode}) and (\ref{eq:hi1})
that the overall water mass balance in the column (box) model is given by
\begin{equation*}
   \sum_{\text{boxes}} m_a(q_v + q_c + q_r) \,=\, M_a - R,
 \end{equation*}
where $M_a$ is the total mass of water within the column (box) at time $t=0$,
and $R$ is the integrated precipitation rate on the ground since $t=0$.
Since our semi-implicit Euler scheme is using in every individual box
the same values $A_1$, $A_2$, $C$, and $E$ in the different equations
and the same inflow and outflow for neighboring boxes, the above identity 
is also maintained for our discrete time evolution, with
%% MHB: changed ``quantities'' to ``sum of \tau''
$R$ being the accumulated sum of $\tau m_a S_{\text{out}}$ of the lowermost box. 

\subsection{Numerical solution of the system 
(\ref{eq:qr12}), (\ref{eq:nr12})}
\label{subsec:qr12nr12}

Solving for $E$ in (\ref{eq:qr12}) and inserting the
corresponding expression into (\ref{eq:Eprime}) 
it follows from (\ref{eq:nr12}) that
\begin{equation*}
  \begin{aligned}
    n_{r,i+1/2}
    &=n_{r,i} - \frac{1}{\overline{m}_r}
    \left(q_{r,i} - q_{r,i+1/2} - 
      \tau S_{\text{out},i+1}(q_{r,i+1/2})\right)\\
    &\qquad- \tau S_{\text{out},i+1}'(n_{r,i+1/2})\\
    &=n_{r,i} - \frac{1}{\overline{m}_r}
    \left(q_{r,i} - q_{r,i+1/2} - \tau s q_{r,i+1/2}\right) 
    - \tau s' n_{r,i+1/2}
  \end{aligned}
\end{equation*}
with
\begin{equation*}
\begin{aligned}
  s = \frac{v_q}{h_i}\,,& &s'=\frac{v_n}{h_i}\,,
\end{aligned}
\end{equation*}
 cf.~(\ref{eq:Sout}).
If we take the average mass of rain drops to be
\begin{equation}
\label{eq:overline-mr}
   \overline{m}_r=q_{r,i}/n_{r,i},
\end{equation}
then it follows that 
$q_{r,i+1/2}$ and $n_{r,i+1/2}$ are linearly coupled, namely
\begin{equation}
\label{eq:qr12-tmp}
   q_{r,i+1/2} = \frac{q_{r,i}}{n_{r,i}}\frac{1+\tau s'}{1+\tau s} \,n_{r,i+1/2}.
\end{equation}
Inserting this into the definition (\ref{eq:Eprime}) of $E'$ we obtain
\begin{equation*}
  \begin{aligned}
  E'&=\frac{n_{r,i}}{q_{r,i}}\, d\rho(q_v-q_{vs})_+
  \left( a_E q_{r,i+1/2}^{\frac{1}{3}}n_{r,i+1/2}^{\frac{2}{3}}
    + b_Eq_{r,i+1/2}^{\frac{1}{2}}n_{r,i+1/2}^{\frac{1}{2}} \right)\\
  &=\lambda n_{r,i+1/2}
\end{aligned}
\end{equation*}
for some (computable) $\lambda>0$. 
It therefore follows from (\ref{eq:nr12}) that $n_{r,i+1/2}$ satisfies
\begin{equation*}
   n_{r,i+1/2} = n_{r,i} - \tau \qty(\lambda n_{r,i+1/2} + s'n_{r,i+1/2}),
 \end{equation*}
 i.e.,
\begin{equation}
\label{eq:nr12-final}
   n_{r,i+1/2} \,=\, \frac{1}{1+\tau\lambda +\tau s'}\,n_{r,i}.
\end{equation}
Inserting (\ref{eq:nr12-final}) into (\ref{eq:qr12-tmp}) we finally obtain
\begin{equation}
\label{eq:qr12-final}
   q_{r,i+1/2} \,=\, \frac{1+\tau s'}{1+\tau s}\frac{1}{1+\tau\lambda+\tau s'}\,
                   q_{r,i}.
\end{equation}
Hence, (\ref{eq:nr12-final}) and (\ref{eq:qr12-final}) are the explicit
solutions of (\ref{eq:qr12}), (\ref{eq:nr12}).
Note that $q_{r,i+1/2}$ and $n_{r,i+1/2}$ remain positive, if $n_{r,i}$ and
$q_{r,i}$ have been positive; see Section~\ref{subsec:open} for the case
when one of the two quantities happens to be zero.

\subsection{Numerical solution of equation (\ref{eq:qci1})}
\label{subsec:Newton}
Given $q_{r,i+1/2}$ and $n_{r,i+1/2}$, the nonlinear equation
(\ref{eq:qci1}) for $q_{c,i+1}$ can be rewritten in the form
\begin{equation*}
   q_{c,i+1} = q_{c,i} + \tau
             \qty(cq_{c,i+1}^\frac{1}{3} - a_1q_{c,i+1}^{2} - a_2 q_{c,i+1})
\end{equation*}
with
\begin{equation*}
\begin{aligned}
   c = d\rho(q_v-q_{vs})n_{c,i}^{\frac{2}{3}}\,,& &
   a_1 = k_1\frac{\rho}{\rho_l}\,,& &
   a_2 = k_2 \pi
  \left(\frac{3}{4\pi \rho_l}\right)^\frac{2}{3}
  v_t\rho q_{r,i+1/2}^\frac{2}{3}n_{r,i+1/2}^\frac{1}{3}\,,
\end{aligned}
\end{equation*}
cf.~(\ref{eq:C}), (\ref{eq:A1}), and (\ref{eq:A2}).
While $a_1$ and $a_2$ are always nonnegative, the sign of $c$ depends on the 
saturation regime: $c$ is positive in supersaturated regimes, and nonpositive
else.

It follows that the nonnegative value of $x=q_{c,i+1}^\frac{1}{3}$ 
is a root of the sixth order polynomial
\begin{equation}
\label{eq:poly}
   p(x) = \tau a_1 x^6 +(1+\tau a_2)x^3 -\tau cx -q_{\text{c},i}\,.
\end{equation}
The first two derivatives of $p$ are given by
\begin{align*}
p'(x) =&\ 6\tau a_1 x^5 + 3(1+\tau a_2)x^2 - \tau c,\\
p''(x)=&\ 30 \tau a_1 x^4 + 6 (1+\tau a_2) x,
\end{align*}
and since $p''(x)>0$ for $x>0$ it follows that the graph of $p$ is strictly convex
for $x\geq 0$. Moreover, $p(0)\leq 0$, and $p(x)\to+\infty$ as $x\to+\infty$.

\begin{figure}%[h]
  \centering
  \includegraphics[height=0.505\linewidth]{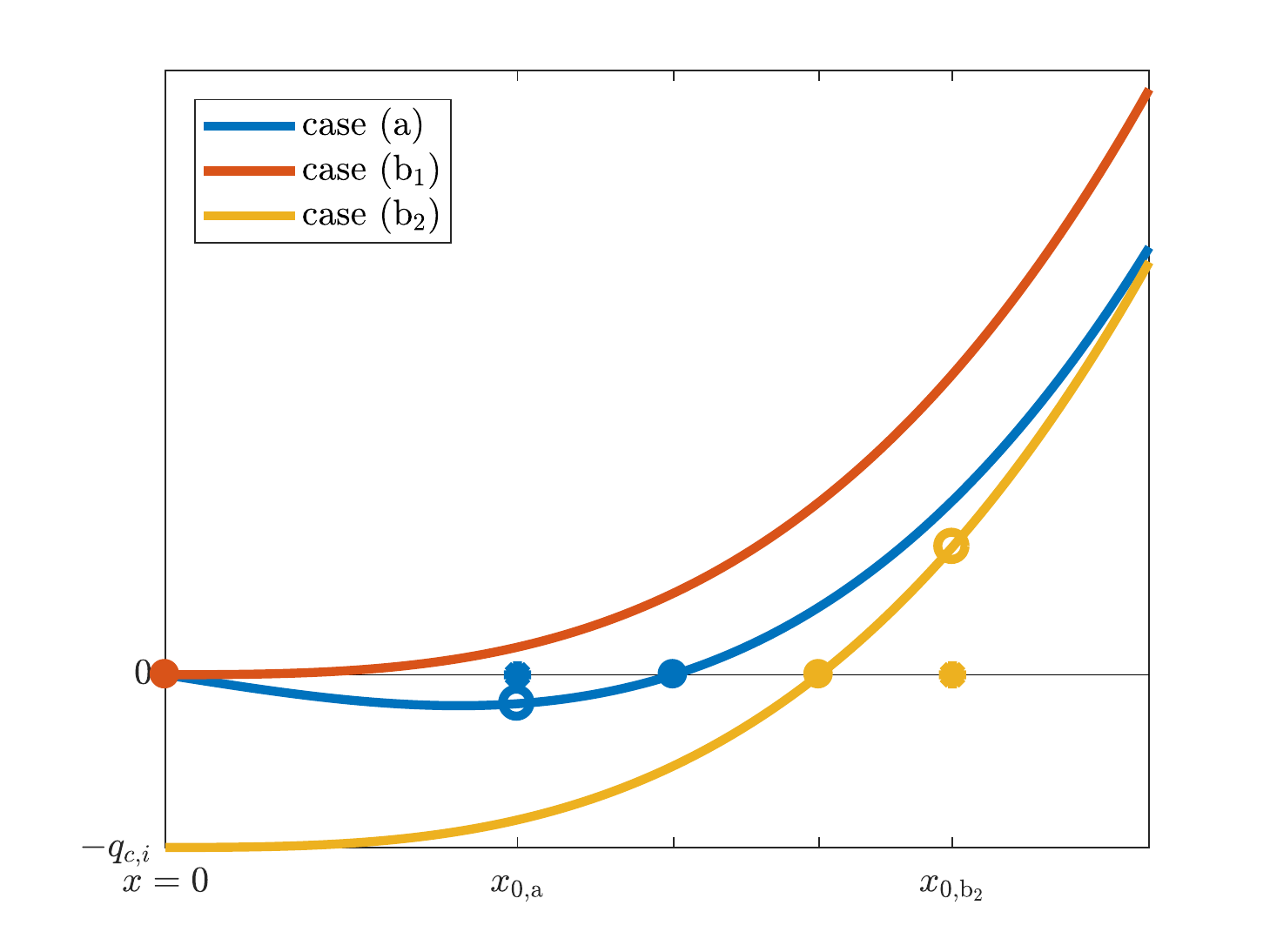}
  \caption{Illustration of the shape of $p$ for Newton's method.}
  \label{Fig:Newton}
\end{figure}

Now we need to distinguish the following cases (see Figure~\ref{Fig:Newton}):
\begin{itemize}
\item[(a)] In the supersaturated regime we have
$p'(0) < 0$, hence $p$ has a unique positive root. 
This root is positive even when $q_{c,i}$ happens to be zero, so that in
this situation the formation of droplets is initiated.
\item[(b)]
In the non-supersaturated regime $c$ is nonpositive and
$p$ is strictly monotonically increasing for $x\geq 0$. 
Accordingly, $p$ has a unique nonnegative root.
\begin{itemize}
\item[(b1)] 
If $q_{c,i}=0$, then $q_{c,i+1}=0$, too; no droplets are generated in this case. 
\item[(b2)] 
If $q_{c,i}>0$, then $p(0)<0$, and hence $q_{c,i+1}$ is strictly positive.
Moreover, $q_{c,i+1}$ is strictly smaller than $q_{c,i}$ in this case, since 
\begin{equation*}
   p(q_{c,i}^\frac{1}{3}) = q_{c,i} - q_{c,i} + 
   \tau(a_1q_{c,i}^2 + a_2q_{c,i}-cq_{c,i}^\frac{1}{3}) > 0,
 \end{equation*}
 because the term in parantheses is strictly positive. 
\end{itemize}
\end{itemize}

Newton's method
\begin{equation*}
  \begin{aligned}
   x_{k+1} = x_k - \frac{p(x_k)}{p'(x_k)},& &k=0,1,2,\dots,
 \end{aligned}
\end{equation*}
is the method of choice for computing the positive root of (\ref{eq:poly})
in the cases (a) and (b2) efficiently.
\begin{itemize}
\item[(a)] 
In the supersaturated regime we recommend to choose
\begin{equation}
\label{eq:x0a}
   x_0 = x_{0,\text{a}} = \min\Bigl\{\left(\frac{\tau c}{3}\right)^\frac{1}{2},
                    \left(\frac{c}{6 a_1}\right)^\frac{1}{5} \Bigr\},
\end{equation}
because $p'(x_0)\geq 0$ in this case, and this guarantees quadratic convergence 
of the Newton iteration. 
\item[(b2)]
In the non-supersaturated regime we suggest to take
\begin{equation}
\label{eq:x0b2}
x_0 = x_{0,\text{b2}} = \left(\frac{q_{c,i}}{\tau a_2+1}\right)^\frac{1}{3},
\end{equation}
because $p(x_0)>0$ for this choice, and again, this guarantees
quadratic convergence of the Newton iteration. 
\end{itemize}
The two initial guesses (\ref{eq:x0a}) and (\ref{eq:x0b2}) are indicated by
circles in Figure~\ref{Fig:Newton}. 

\subsection{Remaining problems}
\label{subsec:open}
Here we address a few peculiarities that may arise in numerical simulations.

\subsubsection{Vanishing rain drop quantities}
Due to round-off it may happen that one of the two variables $q_r$ and $n_r$ 
has become zero at some point, while the other one is still positive.
In that case we naturally set the associated evaporation term $E$ or $E'$,
respectively, to zero, because the corresponding quantity cannot evaporate
as it is not present. It then follows from the respective implicit Euler
equation (\ref{eq:qr12}) or (\ref{eq:nr12}) that this variable
stays zero at the intermediate time step $i+1/2$. 

If $n_{r,i}=0$, then we also conclude from (\ref{eq:E})
that $E=0$, and hence we obtain
\begin{equation*}
  \begin{aligned}
   q_{r,i+1/2} = \frac{1}{1+\tau s}\, q_{r,i}\,,& &n_{r,i+1/2}=0\,;& &
   E = 0\,;& &S_{\text{out}}=sq_{r,i+1/2}\,,& &S_{\text{out}}'=0\,;
 \end{aligned}
\end{equation*}
this is correct, independent of whether $q_{r,i}=0$ as well, or not.

On the other hand, if $q_{r,i}=0$, but $n_{r,i}\neq 0$, then formally,
$1/\overline{m}_r=+\infty$ in (\ref{eq:overline-mr}), so that the 
evaporation rate $E'$ is maximal; cf.~(\ref{eq:Eprime}).
We take this as reasoning to completely ``evaporate'' the remaining number 
concentration in one single time step, and to set $n_{r,i+1}$ and the
associated sedimentation term to zero, i.e.,
\begin{equation*}
  \begin{aligned}
       q_{r,i+1/2} = 0,& &n_{r,i+1/2}=0\,;& &E = 0\,;& &S_{\text{out}} = S_{\text{out}}' = 0.
  \end{aligned}
\end{equation*}
Note that the value of $E'$ is irrelevant for the remaining 
computations in this time step.

\subsubsection{Negative vapour concentrations}
In the supersaturated regime
it may happen that the condensation term becomes larger than the available
vapour mass when the time step is too large. 
Although this is not a very realistic scenario because in this regime
$q_v$ will rarely be sufficently small for this to happen, 
this may lead to a negative value of $q_{v,i+1}$.

One option to cure this problem is to set $q_{v,i+1}=0$ in this case, but this
would result in a gain of water mass. We therefore recommend to reject such a
time step instead, and to repeat the computation with a smaller time step.
 
\subsubsection{Updating the box size}
As mentioned before, cf.~(\ref{eq:hi1}), at the very end of each time step
we need to modify the height of every box to compensate for changes in the
density. In the column model this leads to additional vertical movements
$\Delta w$ of all but the lowermost boxes, and this in turn gives rise to a
corresponding change
\begin{equation}
\label{eq:DeltapT}
\begin{aligned}
  \Delta p = -\gamma \rho \Delta w,& &\Delta T = -\gamma \Delta w,
\end{aligned}
\end{equation}
of pressure and temperature; cf.~(\ref{eq:ode}). 
Different to what has been said right before Section~\ref{subsec:Euler} 
we therefore need to update the pressure and the temperature according 
to (\ref{eq:DeltapT}) right at the beginning of the subsequent time step,
even before the other parameters -- some of which depend on $p$ or $T$ --
are being evaluated.

\section{Numerical simulations}
\label{sec:numtest}

In this section we present some numerical experiments with the new
model.
In Section~\ref{subsec:supersaturation} we demonstrate that our new
model reproduces reasonable levels of supersaturation, while
Section~\ref{subsec:versus_sat_adjust} shows the negative impact of
using saturation adjustment instead. In
Section~\ref{subsec:activation} we compare our cloud activation model
with a more sophisticated scheme, which is based on an effective CCN
distribution. While these first three experiments use the box model, 
the last experiment described in Section~\ref{subsec:showcasing}
considers the full column model and demonstrates that it determines
reasonable amounts of precipitation for three representative updraft
scenarios.

% In Sections~\ref{subsec:supersaturation} and \ref{subsec:activation} we consider a single air parcel with an initial
% height $h=\SI{500}{\meter}$ and in Section~\ref{subsec:showcasing} we consider a column of five air parcels with a total initial height $h=\SI{1000}{\meter}$.
% On the one hand these simulations demonstrate that the 
% numerical method is able to properly determine the time evolution of 
% supersaturation. 
% On the other hand we show that due to our functional relation
% between $n_c$ and $q_c$ our model reproduces recognized features of cloud
% droplet activation.
% % without a sophisticated activation scheme but with our functional relation
% % between $n_c$ and $q_c$. 
% Last, but not least, we demonstrate that the column model determines reasonable
% amounts of precipitation for three representative updraft scenarios. 
% %we show the ability of our new cloud model to produce a reasonable amount of
% %precipitation for each respective scenario.

\subsection{Time evolution of supersaturation}
\label{subsec:supersaturation}

%% MHB: switched to present perfect and reordered first paragraph
As a first test case we have compared our time evolution of
supersaturation with the results published by \citet{korolev_mazin2003}.
To this end we have reduced our model to
the process $C$ of condensation, i.e., collision and sedimentation
processes have been switched off. 
In addition, we have prescribed a fixed number
concentration of cloud droplets. This setup has been used for a direct
comparison with \citet{korolev_mazin2003} for different
sources of supersaturation, i.e., different vertical upward motions,
which drive the adiabatic cooling and thus provide a permanent source for
supersaturation.

\begin{figure}%[h]
  \centering
  \includegraphics[height=0.505\linewidth]{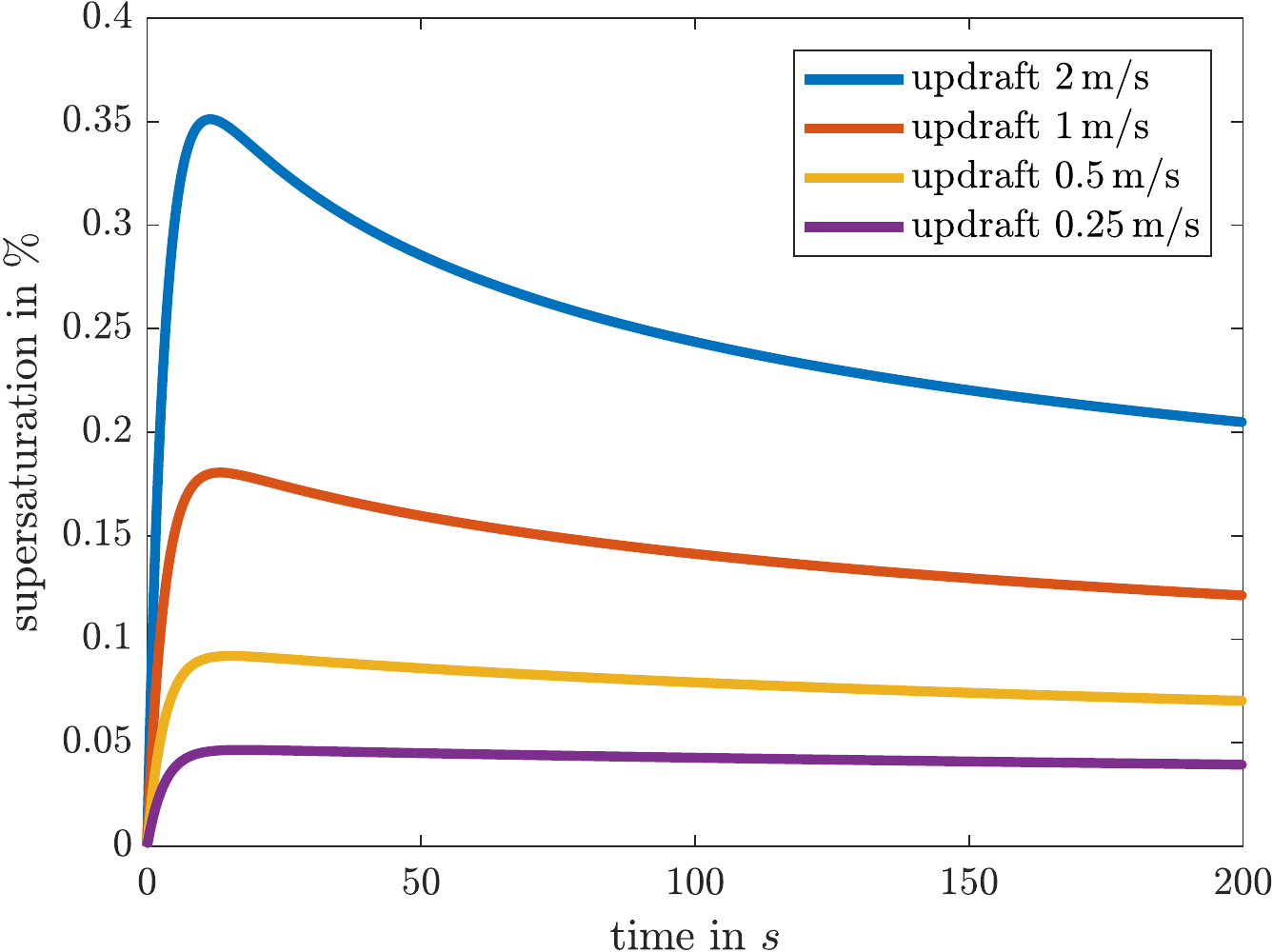}
  \caption{Temporal evolution of the supersaturation ratio $q_v/q_{vs}$ as computed with our cloud model for different updraft velocities. The cloud model reproduces the results of \citet{korolev_mazin2003} almost perfectly.}
  \label{Fig:Korolev}
\end{figure}

The simulation starts in a regime with $q_{vs}-q_v=0$,
%$T(0) = \SI{0}{\celsius}$,
$T(0)=\SI{273.15}{\kelvin}$
and $p(0) = \SI{87000}{\pascal}$ with
$\rho n_c = \SI{2e8}{\per\cubic\meter}$ cloud droplets that have the initial
radius of $\SI{5}{\micro\meter}$. These droplets grow by condensation while their
amount $n_c$ is kept constant. Different vertical velocities are used
for representing different degrees of supersaturation,
i.e., $w\in\{0.25,0.5,1,2\}\,\si{\meter\per\second}$. 
Figure~\ref{Fig:Korolev} shows the results, which agree
almost perfectly with the supersaturation behavior shown by
\citet[][their Figure~1]{korolev_mazin2003}.
Especially, the time evolution of the supersaturation with a peak and
a subsequent decay can be seen. % Note, that the
% semi-implicit numerical scheme has been used here, which is 
% quite stable. 
% Although a small time step $\tau\sim \SI{0.01}{\second}$ has been
% used, the scheme runs stably also for larger time steps; however,
% the peak is then not correctly represented.
Although supersaturation is known to be quite sensitive to changes in
the parameters and therefore also to numerical errors, our
semi-implicit strategy is well suited for the approximation of
supersaturation. % and shows a good agreement with the results from
% \citet{korolev_mazin2003}.

\subsection{The impact of saturation adjustment}
\label{subsec:versus_sat_adjust}

One of the major disadvantages of saturation adjustment shows up in
its influence on cloud buoyancy. The potential density temperature
\begin{equation}
  \label{eq:potential_density_temperature}
  \theta_d:=\theta\left(1+\epsilon_0q_v-q_c\right),~~
  \theta=T\left(\frac{p_0}{p}\right)^\frac{R_a}{c_p}, ~~
  \epsilon_0=\frac{1}{\epsilon}-1\approx 0.608
\end{equation}
% (as defined in
% equation~(\ref{eq:potential_density_temperature}))
in the air parcel and its difference to the corresponding quantity of
the environment determines the buoyancy and thus the time evolution of
vertical velocities. If saturation adjustment is introduced, more
latent heat will be produced, which results in higher updraft
velocities and stronger cooling of the air parcel, and hence to
further condensation, a vicious cycle. In this section we want to
quantify the impact on $\theta_d$ if our model would use saturation
adjustment instead of tolerating supersaturation. This study is similar
in nature to one by \citet{grabowski_jarecka2015}, although the
feedback effect cannot be treated in our simple box model, because our
upward motion is predefined.

% A larger temperature difference due to more
% latent heat release would lead to higher updrafts; this in turn
% results into larger supersaturation and cloud water, feeding back to
% $\theta_d$.

% However, this effect cannot be treated in our simple setups,
% prescribing vertical upward motions without impact of processes in the
% cloud on the driving velocity.
% In the following we investigate
% possible effects of our more physical formulation of condensation in
% contrast to saturation adjustment using our simple model setup. For
% this purpose,
% The major disadvantages, the cutoff at water saturation has, show most
% significantly in the influence on the buoyancy through a back feed on
% the surrounding flow which has not yet been done with this model. 
% Therefore we hint the difference and the influence that an explicitly
% formulated condensation has in comparison to a scheme that uses the
% saturation adjustment method. To this end, 
% we implemented a variation
% of our model which only differs in those parts essential to adjust the
% saturation.
For the implementation of saturation adjustment we have formulated a
complementarity problem for $q_c+q_v$, which is then solved numerically
using a Newton scheme. This replaces the nonlinear
equation~\eqref{eq:qci1}, and thus corresponds to a different
activation scheme.

\begin{figure}[t]
  \centering
  \includegraphics[width=0.5\linewidth]{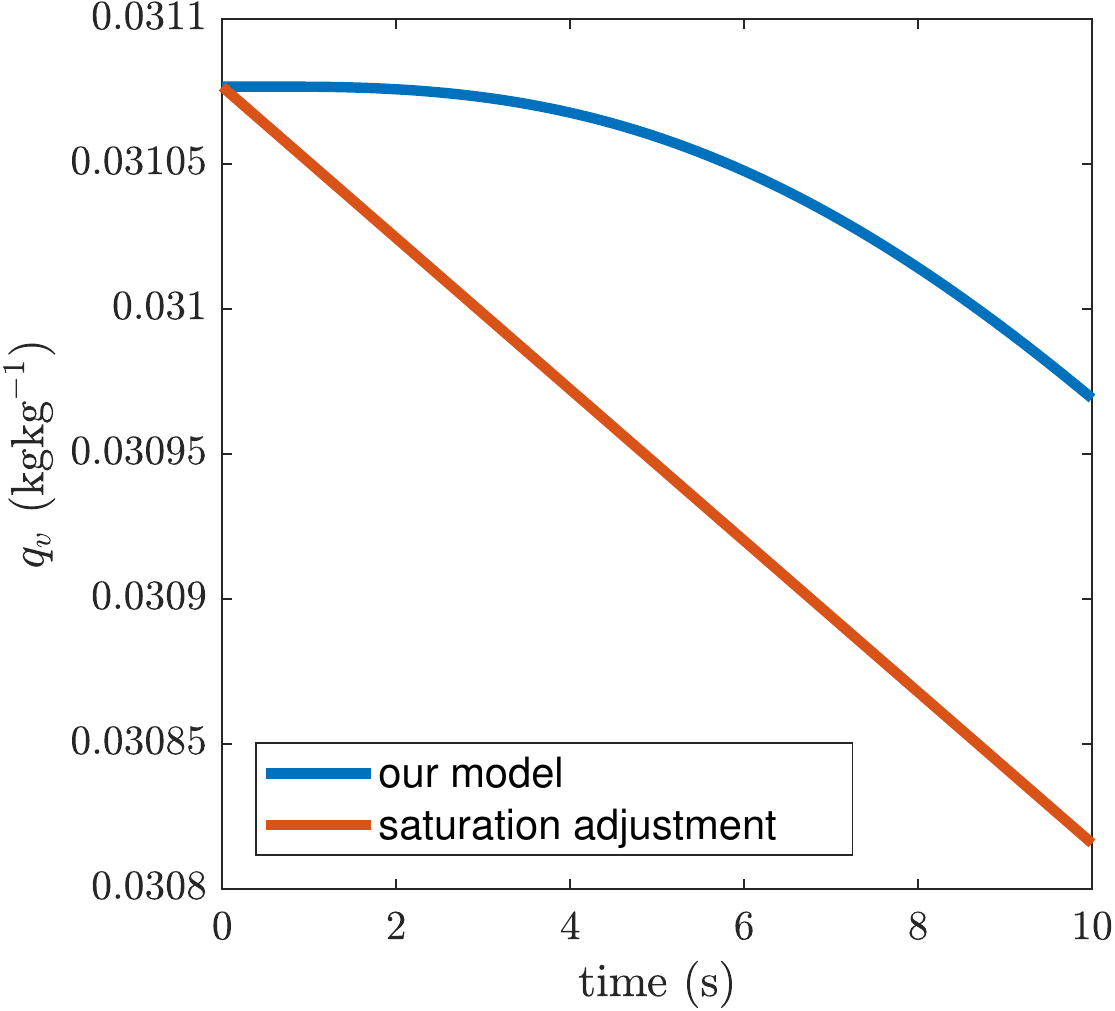}
  \includegraphics[width=0.48\linewidth]{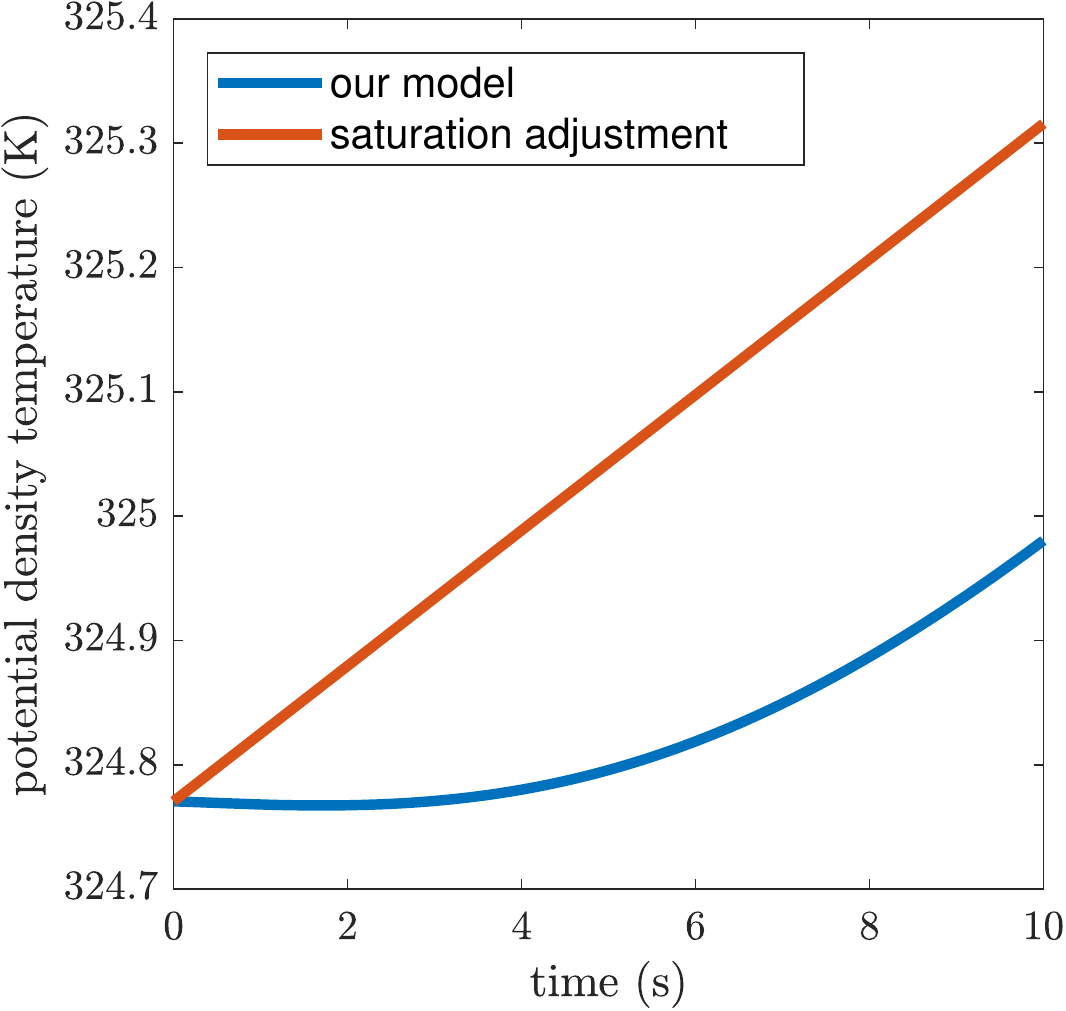}
  \caption{Evolution of water vapour concentration $q_v$ (left) and
    potential density temperature $\theta_d$ (right) for the two schemes.} 
  \label{Fig:SatAdjust}
\end{figure}

% In the study by \citet{grabowski_jarecka2015} the impact of saturation
% adjustment in comparison to a scheme allowing supersaturation is
% investigated. Roughly speaking, it is asumed that within one time
% step, supersaturation of a certain value is reached and the difference
% in the resulting potential density temperatures for both schemes. We
% could reproduce their results in a very good agreement. However, for a
% more detailed insight about the effects of saturation adjustment and
% our more physical representation of condensation,

We have set up an experiment in a maritime environment, i.e., using
$N_\infty=\SI{8e6}{\per\kilogram}$, an initial temperature of
$\SI{303.15}{\kelvin}$, and an initial pressure of
$\SI{85000}{\pascal}$.  We further have prescribed a constant updraft
($w=\SI{10}{\meter\per\second}$) with an air parcel at water
saturation, but without any liquid water. In our model the parcel is
then permanently supersaturated, while the alternative model with saturation adjustment keeps it
at saturation at any time. The time evolution of water vapour
concentration $q_v$ and the potential density temperature $\theta_d$
of the two methods are shown in Figure~\ref{Fig:SatAdjust}. It is
clear that with saturation adjustment more water vapour is
depleted and more latent heat is released as compared to our
model. Therefore, the increase of potential density temperature is
more pronounced for saturation adjustment: After $t\sim \SI{10}{\second}$ the resulting
difference is $\Delta \theta_d\approx
\SI{0.3}{\kelvin}$. Although this seems quite marginal, it will
already introduce a non-negligible additional buoyancy as has been
demonstrated in \citet{grabowski_jarecka2015}. This supports our
modeling decision to develop a cloud scheme without saturation
adjustment. 

\subsection{Activation of cloud droplets}
\label{subsec:activation}

Next we show that 
our strategy of taking the cloud droplet number concentration 
to be a nonlinear function of the droplet mass concentration
%% MHB: changed ``events'' to ``counts''
yields reasonable droplet activation counts, by comparing
our model with a sophisticated activation scheme. 
To this end, we have switched off the processes $A_1$ and $A_2$, 
and have focused on the activation of cloud droplets within
a continental regime (high amount of CCNs) and a maritime environment
(low amount of CCNs).

\subsubsection{Activation schemes}
\label{subsubsec:activation_schemes}

The sophisticated reference model uses a two moment scheme for a single
air parcel, i.e., it explicitely
tracks the number and mass concentrations $n_c$ and $q_c$ of the cloud droplets
together with the evolution of pressure, temperature, and saturation
ratio $\frac{q_v}{q_{vs}}$. The equations for pressure, temperature, and
the condensation process are the same as in \eqref{eq:ode} and
\eqref{eq:C}, respectively. In addition to the condensation process, the
differential equation for $q_c$ includes the change in number concentration 
by assuming a constant mass for a newly activated droplet with radius
\SI{0.5}{\micro\meter}, resembling the choice of $m_0$ in
\eqref{eq:nc_relation}, see also Table~\ref{tab:appendix_model_parameters}.

The equation for the number concentration $n_c$ in the two-moment scheme
is given by
\begin{equation}
  \label{eq:activation_scheme_eq_dnc_dt}
  \dot{n}_c\,=\,\frac{1}{\tau_{\mathrm{act}}}\,\bigl(N_{\mathrm{CCN}}-n_c\bigr)_+
                \frac{(q_v-q_{vs})_+}{q_v-q_{vs}}
\end{equation}
%\begin{equation}
%  \label{eq:activation_scheme_eq_dnc_dt}
%  \dot{n}_c=\max\qty(0,\,\tau_{\mathrm{act}}^{-1}\qty(N_{\mathrm{CCN}}
%   \qty(\frac{q_{v}}{q_{vs}})-n_c))\cdot \frac{(q_v-q_{vs})_+}{q_v-q_{vs}}
%\end{equation}
with an activation timescale $\tau_{\mathrm{act}}=\SI{1}{\second}$ and
a CCN spectrum 
%$N_{\mathrm{CCN}}$ 
$N_{\mathrm{CCN}}$ which depends on the saturation ratio
$q_v/q_{vs}$, 
following \citet{morrison_grabowski2007}.
The last factor in
\eqref{eq:activation_scheme_eq_dnc_dt} ensures that the number concentration can
only change in the supersaturated regime. For the choice of the CCN spectrum 
%$N_{\mathrm{CCN}}$ 
$N_{\mathrm{CCN}}$ there are essentially two possibilities
\citep[see][]{herbert_wacker1998}. The first possibility
is based on a background aerosol particle distribution, resulting
in a detailed, but expensive scheme based on Köhler theory.
The second possibility, as is employed here, is the choice
of an empirical relationship, in particular a power-law
relation (``Twomey spectrum'')
\begin{equation}
  \label{eq:activation_scheme_choice_NCCN}
  N_{\mathrm{CCN}}=C_{\mathrm{CCN}}(q_v/q_{vs}-1)^{\kappa}
\end{equation}
with positive parameters $C_{\mathrm{CCN}}$ and $\kappa$. The choice of
$C_{\mathrm{CCN}}$ encodes the typical background aerosol number (being different
for continental and maritime scenarios) and the exponent $\kappa<1$ adjusts the
steepness of $N_{\mathrm{CCN}}$ near $q_v=q_{vs}$, i.e., the sensitivity of
$N_{\mathrm{CCN}}$ for small levels of supersaturation.
%, in particular if $0<\kappa<1$.

\subsubsection{Maritime and continental scenarios}
\label{subsubsec:scenarios}

In the following, we consider a continental and a maritime case to
compare our cloud model with the detailed two moment scheme described
in the preceding section. In both cases we consider a single air parcel,
ascending with \SI{2}{\meter\per\second} with initial pressure
\SI{87000}{\pascal} and temperature \SI{273.15}{\kelvin}.
%ps war es wirklich 273.16K???
Since we want to compare the activation of cloud droplets, we assume
an initial humidity $q_v=q_{vs}$, i.e. the air parcel is initially
at saturation. The time step in both models is $\tau=\SI{0.01}{\second}$.
As explained in Section~\ref{subsubsec:particle_formation}, in our
new cloud model the number of cloud droplets is tied to the mass
concentration $q_c$ through \eqref{eq:nc_relation}. Therefore,
we choose the initial number concentration for the two moment
scheme according to this relation and assume in both models
$q_c(0)=\SI{e-10}{\kilogram\per\kilogram}$, i.e. an essentially
cloud free case. The choices for $N_0$ and $m_0$ are given in
Table~\ref{tab:appendix_model_parameters}.

{\color{red} }

\begin{figure}
	\centering
	\includegraphics[height=0.505\linewidth]{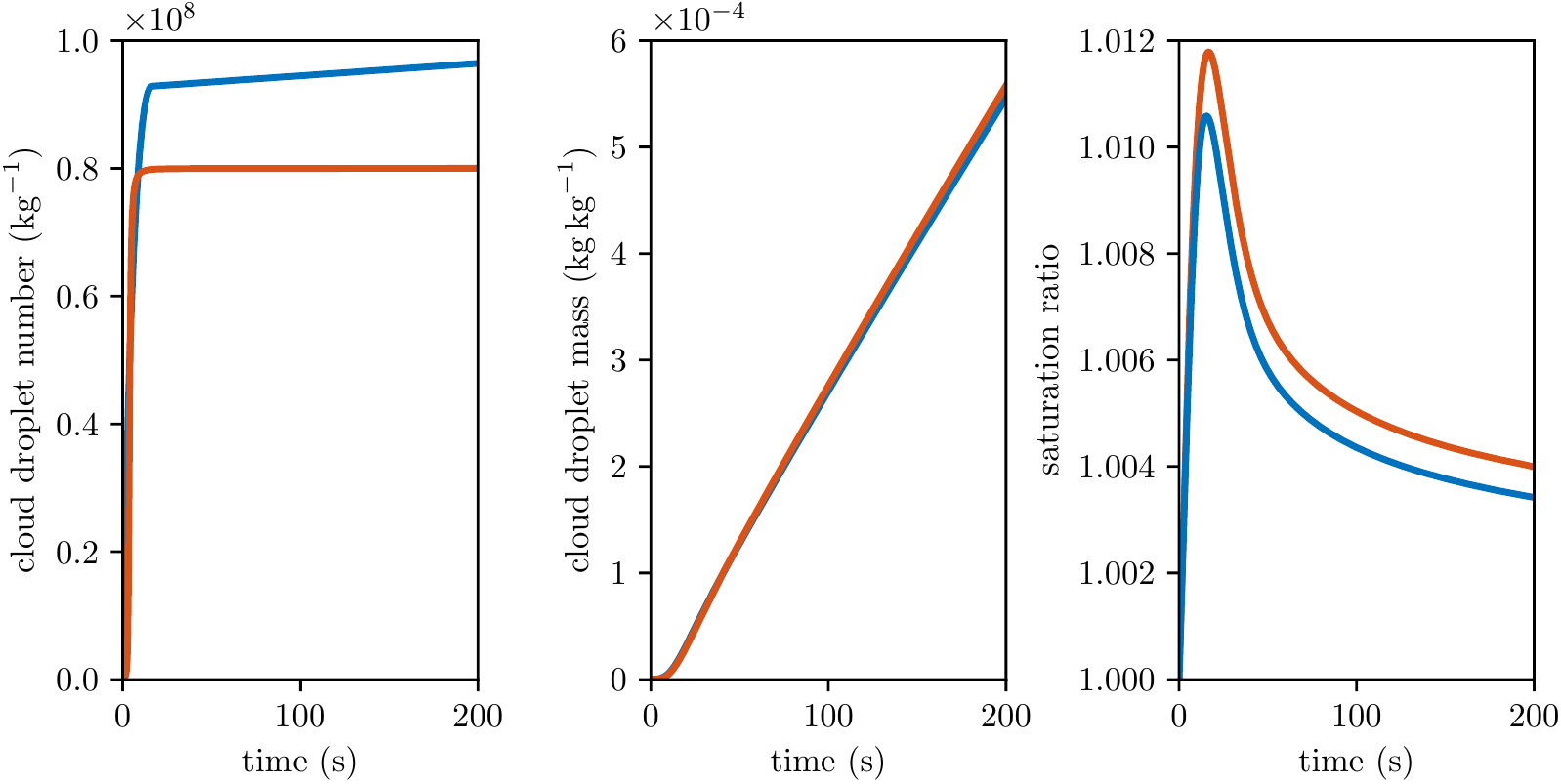}
	\caption{Maritime test case: our scheme (red) versus reference scheme (blue)}
	\label{fig:comp_maritime}
\end{figure}

For the maritime case, the parameter choices for the CCN-spectrum
\eqref{eq:activation_scheme_choice_NCCN} are given by
% Einheiten: $C_{\mathrm{CCN}}=\SI{1e9}{\per\cubic\meter}$
$C_{\mathrm{CCN}}=\SI{9e8}{\per\kilogram}$ and $\kappa=\frac{1}{2}$,
so that at \SI{1}{\percent} supersaturation the number concentration
of cloud droplets corresponds to the figure of
\SI{100}{\per\cubic\centi\meter} tabulated in
\citet[][]{pruppacher_klett2010}.  The free parameter $N_{\infty}$ in
\eqref{eq:nc_relation} is selected from the interval
$\SI{6e7}{\per\kilogram}\le N_{\infty}\le
\SI{1e8}{\per\kilogram}$. % Note that we can switch between
% concentrations and densities with the help of the air density $\rho$.
Figure~\ref{fig:comp_maritime} shows the saturation ratio $q_v/q_{vs}$
(right panel), the mass concentration (middle panel), and the number
concentration (left panel) for the two moment scheme (blue curves)
and the new cloud model (red curves) with parameter
$N_{\infty}=\SI{8e7}{\per\kilogram}$.
% Figure~\ref{Fig:Comp_sat_maritime} shows the saturation ratio
% $q_v/q_{vs}$, Figure~\ref{Fig:Comp_mass_maritime} the mass
% concentration, and Figure~\ref{Fig:Comp_number_maritime} the number
% concentration for the new cloud model (red curves) and the two moment
% scheme (blue curves).
%% Referee: be more precise! (item 9 on his/her list)
We observe a good agreement of the supersaturation ratio and the cloud
droplet mass concentration for both schemes, see also 
Table~\ref{tab:quantification_maritim}.
%in all parameters to the results of the two moment scheme.

\begin{figure}[h]
	\centering
	\includegraphics[height=0.505\linewidth]{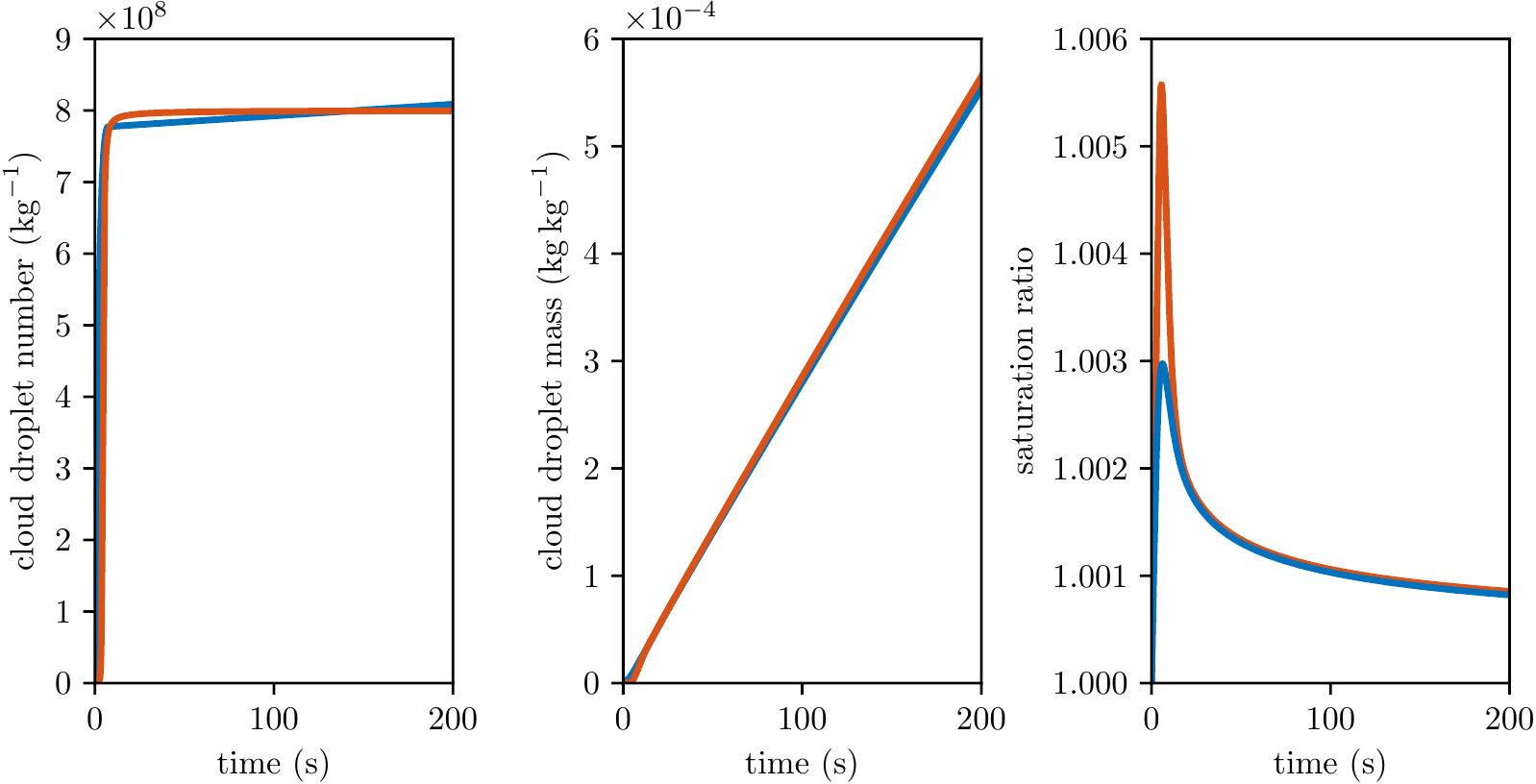}
	\caption{Continental test case: our scheme (red) versus reference scheme (blue)}
	\label{fig:comp_continental}
\end{figure}

For the continental case, we choose the parameters
%$C_{\mathrm{CCN}}=\SI{5.2044e9}{\per\cubic\meter}$
$C_{\mathrm{CCN}}=\SI{4.69e9}{\per\kilogram}$ and $\kappa=\num{0.308}$
in \eqref{eq:activation_scheme_choice_NCCN}, so that at
\SI{1}{\percent} supersaturation the number concentration of cloud
droplets is corresponding to \SI{1260}{\per\cubic\centi\meter}
\citep[as given in][]{seifert_beheng2006}.  Here we vary $N_{\infty}$
between $\SI{6e8}{\per\kilogram}$ and
$\SI{1e9}{\per\kilogram}$ because the number of CCN is almost always one order of magnitude larger
over land than over sea \citep{pruppacher_klett2010}. Figure~\ref{fig:comp_continental} shows the
corresponding results with the choice
$N_{\infty}=\SI{8e8}{\per\kilogram}$: saturation ratio (right
panel), mass concentration (middle panel), and  number
concentration (left panel).
% in Figure~\ref{Fig:Comp_sat_continental} for the saturation ratio,
% Figure~\ref{Fig:Comp_mass_continental} for the mass concentration, and
% Figure~\ref{Fig:Comp_number_continental} for the number concentration
% respectively.
Although our cloud model slightly overestimates the
saturation ratio, we  again observe a
%% Referee: be more precise!
good agreement for the mass and number concentrations, see
Table~\ref{tab:quantification_continental}.
%Note that in the continental case we have chosen the free parameter
%$N_\infty$ one order of magnitude larger than
%in the maritime case, 
%%being consistent with the fact that over
%because the number of CCN is almost always one order of magnitude larger
%over land than over sea \citep{pruppacher_klett2010}.

It is impressive that the fixed coupling of mass and number of the cloud
droplets 
%$q_c$ and the number of cloud droplets $n_c$ 
yields such a good agreement of our numerical results with those of
the much more detailed two moment scheme, where the number and mass
concentrations evolve independently.  In our model, we only have
adjusted the parameter $N_{\infty}$ in \eqref{eq:nc_relation} to the
given regime (maritime or continental).
% and we end up with the results shown above.
Keeping this in mind, the agreement of the saturation ratio between
our cloud model and the two moment scheme is surprisingly good, given
the sensitivity of the saturation ratio with respect to changes in the
modeling of the condensation process and the mass concentration.

% As already mentioned, the
% saturation ratio $q_v/q_{vs}$ is quite sensitive to changes in
% the condensation process through the release of latent heat. Consequently,
% the saturation ratio is also sensitive to the mass concentration.
% Keeping such relationships in mind, the agreement of the saturation ratio
% between our cloud model and the two moment scheme is surprisingly good.

%ps vier nachkommastellen
\begin{table}
	\centering
	\begin{tabular}{ccc}
		$N_\infty[\si{\per\kilogram}]$ & {\rm saturation ratio}
                & 
                                                    $q_c$ \\ 
		\hline
		% $6\cdot10^7$& 1.013481464204793 & 0.016727341750881 \\
		% $8\cdot10^7$& 1.011788371611779 & 0.019924698209468 \\
		% $10\cdot10^7$& 1.010636968740548 & 0.023988176841826 \\
		$\phantom{1}6\cdot10^7$&  $5.61\cdot10^{-4}$ & $0.0167$ \\
		$\phantom{1}8\cdot10^7$&  $3.88\cdot10^{-4}$ & $0.0199$ \\
		$10\cdot10^7$& $3.58\cdot10^{-4}$ & $0.0240$ \\
	\end{tabular}
	\caption{Relative root mean square errors for the maritime
          case, depending on the choice of $N_\infty$.}
	\label{tab:quantification_maritim}
      \end{table}

% Continental environment:

% \begin{itemize}
% \item activation scheme: 
%   \begin{enumerate}
%   \item CCN spectrum choice:
%     $\text{C}=\SI{1.26e9}{\per\cubic\meter} = \SI{1.26e3}{\per\cubic\centi\meter}$ and
%     $k=0.308$ from seifert\_axel.pdf table 3.3 row 6 Texas,USA from
%     Khain et al. (2001)
%   \item $N_l = \rho n_c(q_{c,0})$ which depends on the choices of the Parameters in $n_c$
%   \item $q_{c,0}= \num{e-10}$
%   \end{enumerate}
% \item cloud model:
%   \begin{enumerate}
%   \item $N_\infty =\num{8e8}$ (wrong case: $N_\infty =\num{2e8}$)
%   \item $m_0\ \hat{=}\ \SI{0.5e-6}{\meter}$ radius
%   \item $q_{c,0}= \num{e-10}$
%   \end{enumerate}
% \end{itemize}
\begin{table}
	\centering
	\begin{tabular}{ccc}
		$N_\infty[\si{\per\kilogram}]$ & {\rm saturation ratio}
                & 
                                                    $q_c$ \\ 
		\hline
		% $6\cdot10^8$& 1.005811764662711 & 0.018861705780918 \\
		% $8\cdot10^8$& 1.005573542617414 & 0.019829649120500 \\
		% $10\cdot10^8$& 1.005438145598398 & 0.020556559957652\\
		$\phantom{1}6\cdot10^8$&  $2.09\cdot10^{-3}$ & $0.0189$ \\
		$\phantom{1}8\cdot10^8$&  $8.00\cdot10^{-4}$ & $0.0198$ \\
		$10\cdot10^8$& $1.20\cdot10^{-4}$ & $0.0206$ \\
	\end{tabular}
	\caption{Relative root mean square errors for the continental
          case, depending on the choice of $N_\infty$.}
	\label{tab:quantification_continental}
\end{table}

% \begin{figure}%[h]
%   \centering
%   \includegraphics[width=0.8\linewidth]{Vergl_Saettigung_Kontinental.eps}
%   \caption{Saturation ratio $q_v/q_{vs}$ for the continental
%     test case. Red curve: the new cloud model; blue curve: the two
%     moment scheme.} 
%   \label{Fig:Comp_sat_continental}
% \end{figure}
% \begin{figure}%[h]
%   \centering
%   \includegraphics[width=0.8\linewidth]{Vergl_Masse_Kontinental.eps}
%   \caption{Cloud droplet mass concentration $q_c$ for the continental
%     test case. Red curve: the new cloud model; blue curve: the two
%     moment scheme.} 
%   \label{Fig:Comp_mass_continental}
% \end{figure}
% \begin{figure}%[h]
%   \centering
% %  \includegraphics[width=0.8\linewidth]{Vergl_Anzahl_Kontinental.eps}
%   \includegraphics[width=0.8\linewidth]{Anzahl_Kontinental_prokg.eps}
%   \caption{Cloud droplet number concentration for the continental test
%     case. Red curve: the new cloud model; blue curve: the two moment
%     scheme.} 
%   \label{Fig:Comp_number_continental}
% \end{figure}

%\begin{figure}[h]
%  \centering
%  \includegraphics[width=0.34\linewidth,height=0.48\linewidth]
%  {Vergl_Saettigung_Kontinental.eps}
%  \includegraphics[width=0.32\linewidth,height=0.48\linewidth]
%  {Vergl_Masse_Kontinental.eps}
%  \includegraphics[width=0.32\linewidth,height=0.48\linewidth]
%  {Anzahl_Kontinental_prokg.eps}
%  \caption{Continental test case: Numerical results}
%  \label{fig:comp_continental}
%\end{figure}

% \todo{hier noch genaue Beschreibung der figures}
% We calculated the potential density temperature, which is defined as
% follows:

%% MHB: changed section title and reformulated some of the text of this section
\subsection{The full column model: Three updraft scenarios}
\label{subsec:showcasing}

\begin{figure}%[h]
	\centering
	\includegraphics[height=0.505\linewidth]{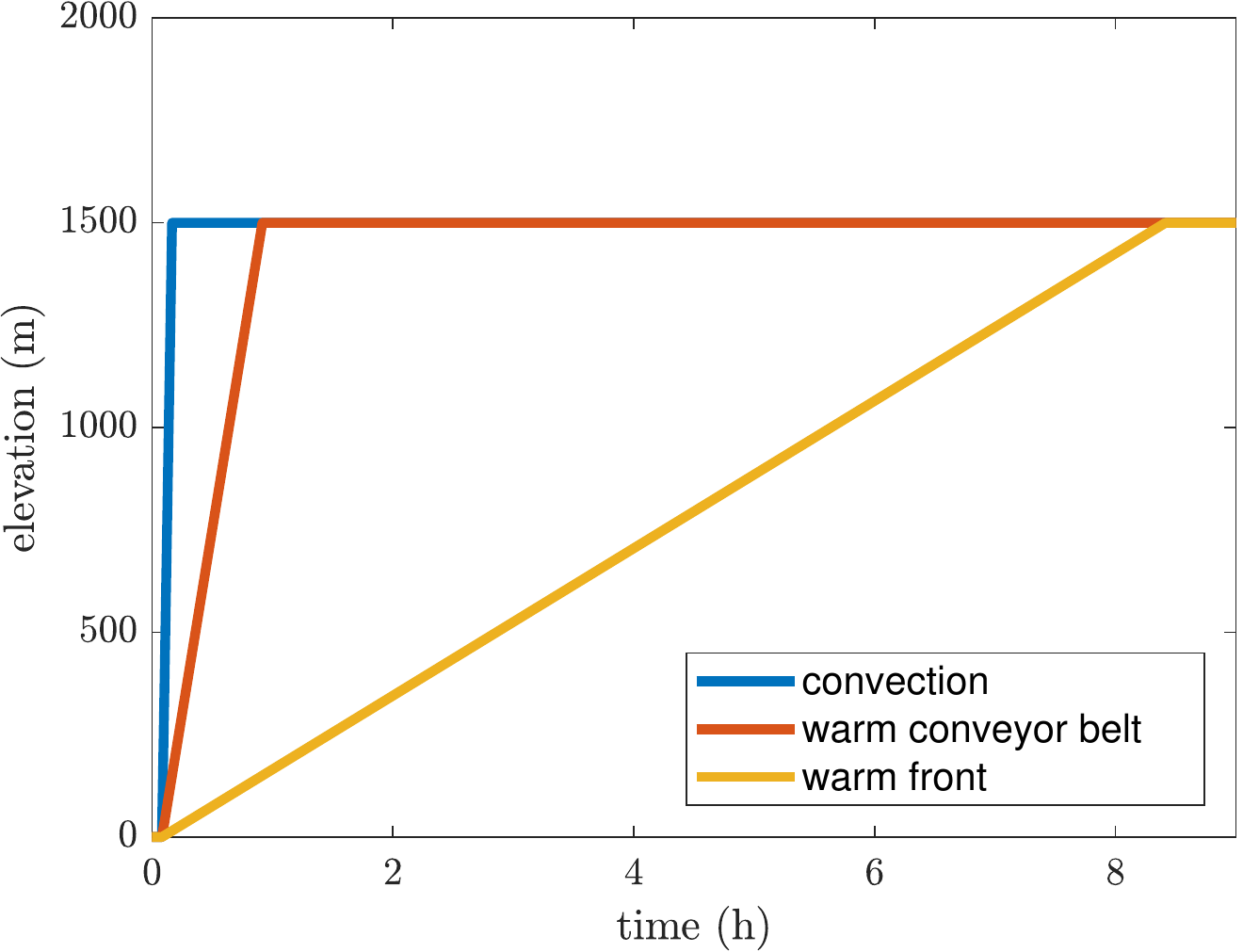}
	\caption{Ascend of the lowermost air parcel of the column over
          time for the three scenarios.} 
	\label{Fig:ascent_profiles}
\end{figure}

Finally we have run the full column model for the following three
updraft scenarios: a warm front with small vertical velocity
$\SI{0.05}{\meter\per\second}$, a warm conveyor belt with moderate
vertical velocity $\SI{0.5}{\meter\per\second}$, and a convective
event with large vertical velocity $\SI{5}{\meter\per\second}$.
%representative average constant vertical wind velocities
%$\SI{0.05}{\meter\per\second}$, $\SI{0.5}{\meter\per\second}$, and
%$\SI{5}{\meter\per\second}$, respectively.  
In each case we have assumed that the column does not ascend in the first
$\SI{5}{\minute}$; thereafter the vertical velocity has been set to the
respective value of this scenario until the bottom of the column has
reached a height of $\SI{1500}{\meter}$, where the vertical velocity has been
set to zero again, see Figure~\ref{Fig:ascent_profiles}.
The column consists of five air parcels, each with an initial height
of \SI{200}{\meter}.  For each scenario, the initial conditions as
well as the saturation profiles are identical: The initial temperature
and pressure for the lowermost air parcel are $\SI{300}{\kelvin}$ and
\SI{101325}{\pascal}, respectively; for the upper air parcels the
temperature and pressure have been initialized according to the adiabatic
lapse rates. The lower three air parcels are initially subsaturated
with \SI{40}{\percent} water saturation, while the upper two air
parcels have initially \SI{80}{\percent} saturation.  With this setup,
the temperature in the whole column never falls below
\SI{275}{\kelvin} during our simulation, hence there occurs no ice, so that our cloud scheme is a reasonable model.

% To compare the results we start the elevation of the columns bottom
% after $\SI{5}{\minute}$ of simulation and then give each column the
% time to elevate to a height of $\SI{1500}{\meter}$ once this height
% is reached we keep the columns bottom at this height for the rest of
% the experiment. For each phenomenon we simulate $\SI{9}{\hour}$ and
% the resulting rain out of the columns bottom can be seen in figure
% \ref{Fig:3_ascent_cases}. Each column is set up with the same
% initial values and saturation profile. The latter is chosen in a way
% that the top third of the column is relatively moist with an initial
% saturation of $80 \%$ while the lower two thirds of the column are
% relatively dry with an initial saturation of $40\%$. The columns are
% initialized with a height of $\SI{1}{\kilo\meter}$ and partitioned
% into 5 equally sized boxes. The lower most box has the initial
% temperature $\SI{300}{\kelvin}$ and the pressure $p_*$ while the
% boxes above are initialized with the adiabatic temperature and
% pressure gradients and their respective heights. Note that the
% temperature falls at no time below $\SI{275}{\kelvin}$ in any point
% of the columns and that the columns height may change over time.

\begin{figure}%[h]
	\centering
	\includegraphics[height=0.505\linewidth]{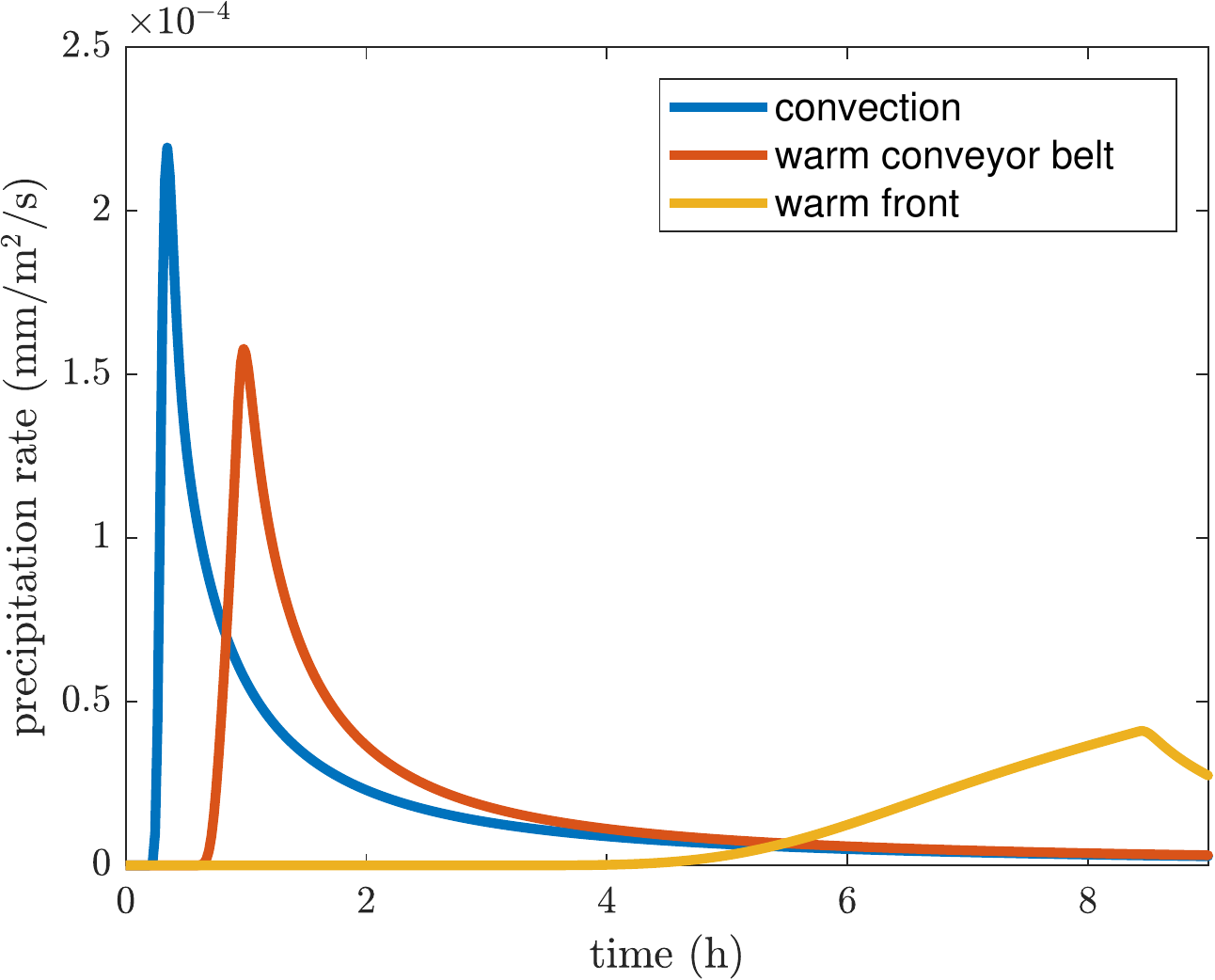}
	\caption{Precipitation rates as a
          function of time.} 
	\label{Fig:3_ascent_cases}
\end{figure}

%% MHB: replaced ``amount of rain, falling out of the column'' 
%%      by ``precipitation rates''
Figure~\ref{Fig:3_ascent_cases} shows the corresponding precipitation rates
%rain, falling out of the column 
as a function of time. The different amounts of rain for
the three cases is realistic: The convective event
produces the largest rainfall in a short time, being in-line with the
large updraft velocity. In contrast, the case of a warm front with
very small updraft velocity produces after about \SI{4}{\hour} only
light but steady rainfall.  Rainfall for the warm conveyor belt is
more similar to the convective case, but with an initial delay of
roughly half an hour, and also the amplitude is smaller.  The relatively
small total precipitation rates in all cases are due to the
subsaturated lower three air parcels of the column. These air parcels
remain cloud free during the whole simulation and do not produce
any rain at all, while the upper two air parcels are cloudy and produce all of the
rain seen in Figure~\ref{Fig:3_ascent_cases}.

% As one can expect the occurrence of precipitation in figure
% \ref{Fig:3_ascent_cases} fits to the respective ascents in figure
% \ref{Fig:ascent_profiles} where the rainfall sets in when the
% saturation level is reached which is fastest for the convection
% event and slowest for the warm front event. Nonetheless the total
% amount of rain is similar in all three events. Since the lower two
% thirds of the column do not reach saturation level during the ascent
% due to their low initial moisture the air parcel producing rain is
% initially only $\SI{400}{\meter}$ high and the small amount of rain
% is therefore reasonable.

\begin{figure}%[h]
	\centering
	\includegraphics[height=0.505\linewidth]{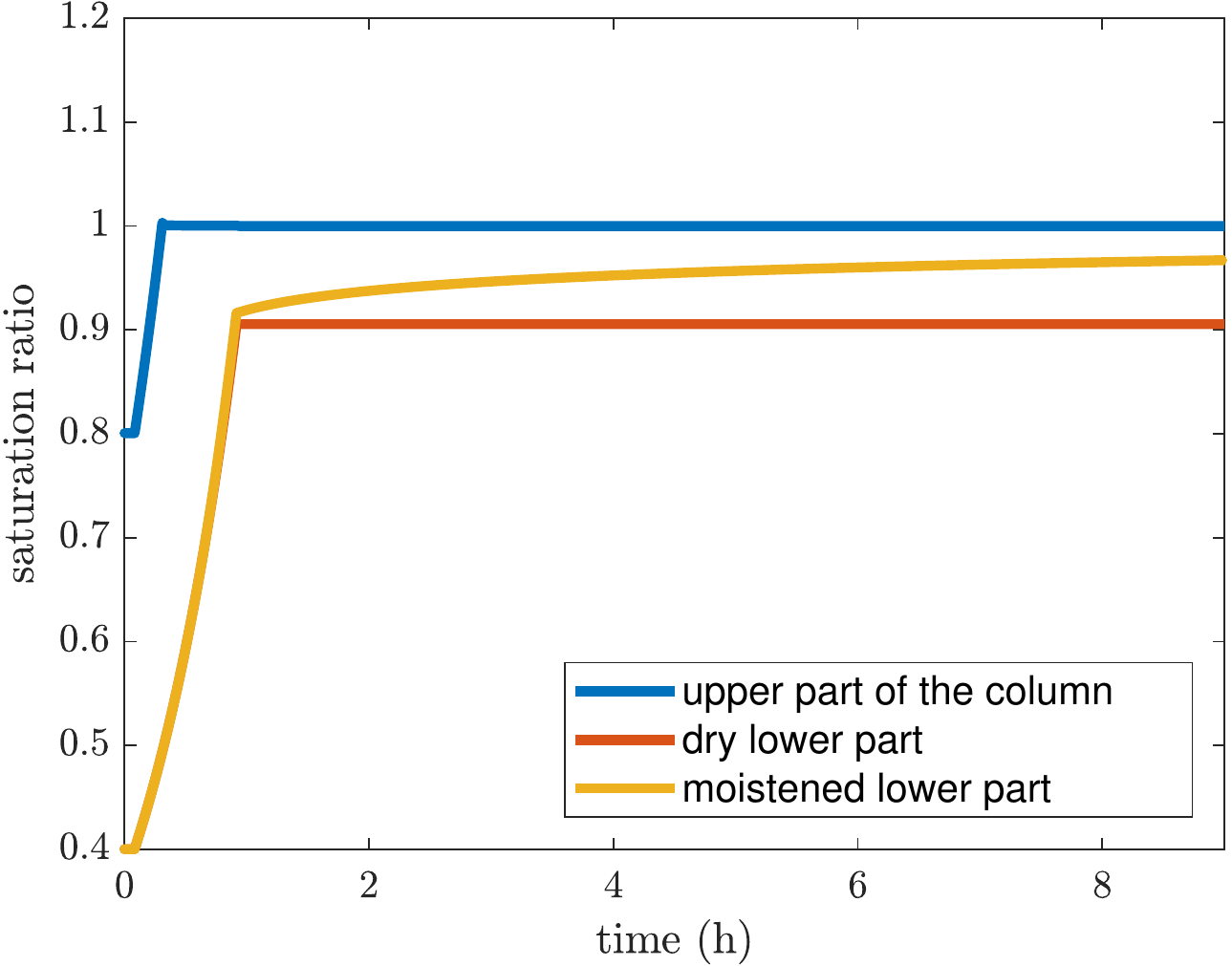}
	\caption{Saturation ratio of the air parcels within the column
          for the warm conveyor belt. Blue curve: upper two air
          parcels of the column; yellow curve: lower part of the
          column; red curve: lower part of the column without the
          moisture of the evaporating rain.} 
	\label{Fig:moistening}
\end{figure}

As already mentioned, the lower part of the column is relatively dry
and gets moistened by the evaporation of rain, falling down from the cloudy
upper part. This effect is illustrated in Figure~\ref{Fig:moistening}
for the case of the warm conveyor belt. The blue line shows the
temporal evolution of the saturation ratio at the bottom of the upper cloudy
part of the column. Due to the updraft, this part gets supersaturated
after roughly \SI{20}{\minute} and the cloud forms. The rainfall
starts after \SI{40}{\minute} and reaches its maximum at roughly
\SI{60}{\minute}, see the red curve in
Figure~\ref{Fig:3_ascent_cases}.  This indicates the delay in the
production of the rain due to the autoconversion process.  After
falling out of the cloudy part of the column, the rain falls through
the subsaturated lower part of the column and evaporates.  To
illustrate the moistening of the lower part we have added the yellow
and the red curve in Figure~\ref{Fig:moistening}: While the yellow
curve displays the actual saturation ratio in the lower part of the
column, the red curve shows the corresponding data, if the evaporation of
rain were switched off. Consequently, the difference of these curves
indicates the moistening of the lower part of the column.

In Sections~\ref{subsec:column_model} and \ref{subsec:mass} we have
discussed the conservation of the total water mass in our cloud model,
when one accounts for the amount of rain falling out of the column.
The absolute loss of total water mass that we have observed in our
simulations are \SI{2.19e-14}{\kilogram\per\kilogram} for the warm front,
\SI{3.97e-15}{\kilogram\per\kilogram} for the warm conveyor belt, and
\SI{1.57e-15}{\kilogram\per\kilogram} for the convective case. These
values confirm the conservation of mass numerically.

% We have also compared our results with the more complicated activation model
% used as a reference in Section~\ref{subsec:activation}.
% We refer to 
% %To quantify the agreement with the more complicated activation model
% %used as a reference in Section~\ref{subsec:activation} we refer to
% Tables \ref{tab:quantification_maritim} and
% \ref{tab:quantification_continental} concerning the relative root mean
% square error of $q_c$ depending on the allegedly
% most influential parameter $N_\infty$.
% One can see that a smaller choice of $N_\infty$
% would even give a better fit of $q_c$ in both cases.

% As mentioned before the lower parts of the columns do not reach the
% saturation level. In figure \ref{Fig:moistening} we can see that for
% the warm conveyor event which follows the red elevation profile in
% figure \ref{Fig:ascent_profiles}. Nevertheless we can see how the
% highest box of the dry lower parts (red curve) gets moistened be the
% precipitation that falls in from the lowest box of the moist upper
% parts (blue curve) of the column after the ascent is completed due
% to evaporation. For comparison we also show an ascent where
% prohibited evaporation in the lower parts (yellow curve).

% Finally we apply the column model to a few vertical wind profiles to
% show the numerics at work. The column with a height of 1-2 km
% initialized with a couple of saturation profiles performs some
% elevations that we consider representative for some particular
% phenomena of ascending air. A saturation profile that allows
% condensation in upper parts of the column

%% MHB: rephrased some sentences
\section{Summary and conclusions}
\label{sec:conclusion}

In this paper we have developed a new model for warm clouds based on
averaged mass and number concentrations of cloud
droplets and rain drops. The model consists of a one-and-a-half moment
scheme with differential equations for the mass concentration of cloud
droplets, and the mass and number concentrations for rain drops.  
 To account for a realisitic activation of cloud
droplets
we do not use a differential equation for the droplet number
concentration, but we couple the droplet number and mass concentrations
directly with a nonlinear functional relation instead.
%between $n_c$ and $q_c$. Thus, we do
%not use a differential equation for $n_c$ but mimic the activation of
%droplets with this approach. 
Growth and evaporation of cloud droplets are realized in 
%a physically adequate way, allowing
such a way that a certain level of supersaturation with respect to
water is tolerated, i.e., we do not apply any sort of saturation
adjustment. The collision terms are given by
rates, which are nonlinear in the mass concentrations, similar to
previous model developments. The sedimentation of rain drops is formulated 
%also included, both 
for a zero-dimensional box model and for a vertical
column model.

%A consistent and adequate numerical scheme 
For the implementation of
this model we propose a consistent numerical scheme, which is semi-implicit,
i.e., some terms are treated explicitly while others are solved for implicitly.
% treated parts and implicitly solved parts, respectively. 
Implicit solvers are necessary, for example, 
to activate cloud droplets, because for zero initial conditions 
explicit solvers will always stick to the nonphysical trivial solution.  
The corresponding implicit equation is
solved using Newton’s method, assuring a quadratic convergence.
%, as shown.  
The implementation is proven to return nonnegative
concentrations only, and to preserve the air and water mass balance.
% This is a non-trivial issue and note that at
%least to our knowledge no mass-conserving Lagrangian column models are
%available in literature.
 
The model has been successfully tested on idealized model setups. We
have compared our numerical results with reference data for the time
evolution of supersaturation presented in
\citet{korolev_mazin2003}. In addition, we have demonstrated the negative effect
of saturation adjustment as compared to our model formulation.  We
have also implemented and run a more sophisticated explicit droplet activation
scheme
% to show the performance of
%the new numerical scheme. In this slightly adjusted model setup (no
%activation, $n_c=\text{const}$) the agreement between the models is
%very good. 
for two different meteorological scenarios (continental vs.\ maritime
environment): In both cases we have seen a very good agreement of our scheme.
% we have also seen a very good agreement of our implicit
%droplet activation scheme and of a more sophisticated approach using
%an explicit activation scheme.
%Actually, our simple activation
%scheme behave very meaningful, showing the increase of cloud number
%concentration durign the activation process. 
%We remark that for all three key variables (supersaturation, mass/number
%concentration of cloud droplets) the time evolution is very similar to
%the sophisticated model.
%Finally,
Finally, we have also shown that our column model determines
meaningful rain formation pathways and precipitation data for
different updraft scenarios, such as slow
%simulationcorrectly simulates able to simulate 
%consistently a column of air parcels, where all physical processes 
%described in Section~\ref{sec:model_description}
%are active. We carried out numerical simulations with a
%column consisting of five air parcels, ascending with different
%vertical velocities, matching typical conditions for a slow
frontal updraft, warm conveyor belts, and convective events, respectively.

\section{Acknowledgement}
Nikolas Porz acknowledges support of the Transregional Collaborative
  Research Center SFB/TRR 165 ``Waves to Weather'', funded by the
  Deutsche Forschungsgemeinschaft (DFG), within the subproject
  ``Identification of robust cloud patterns via inverse methods''.
  Manuel Baumgartner acknowledges support of the Deutsche
  Forschungsgemeinschaft (DFG) within the project ``Enabling
  Performance Engineering in Hesse and Rhineland-Palatinate'' (grant
  number 320898076).  We thank Philipp Reutter for interesting
%% MHB: deleted ``possible justifications''
  discussions of our droplet activation process.
%%%%%%%%%%%%%%%%%%%%%%%%%%%%%%%%%%%%%%%%%%%%%%%%%%%%%%%%%%%%%%%%%%%%%%%%%%%%%%%%
%%%%%%%%%%%%%%%%%%%%%%%%%%%%%%%%%%%%%%%%%%%%%%%%%%%%%%%%%%%%%%%%%%%%%%%%%%%%%%%%

%%%%%%%%%%%%%%%%%%%%%%%%%%%%%%%%%%%%%%%%%%%%%%%%%%%%%%%%%%%%%%%%%%%%%%%%%%%%%%%%
%%%%%%%%%%%%%%%%%%%%%%%%%%%%%%%%%%%%%%%%%%%%%%%%%%%%%%%%%%%%%%%%%%%%%%%%%%%%%%%%
\appendix
%%%%%%%%%%%%%%%%%%%%%%%%%%%%%%%%%%%%%%%%%%%%%%%%%%%%%%%%%%%%%%%%%%%%%%%%%%%%%%%%
%%%%%%%%%%%%%%%%%%%%%%%%%%%%%%%%%%%%%%%%%%%%%%%%%%%%%%%%%%%%%%%%%%%%%%%%%%%%%%%%

\renewcommand{\thesection}{\Alph{section}}
\setcounter{section}{0}
\section{Technical details of the overall algorithm}
%\label{appendixA:flowchart}
\label{appendixB:constants}

\begin{figure}
	\centering
	\includegraphics[height=0.93\textheight]{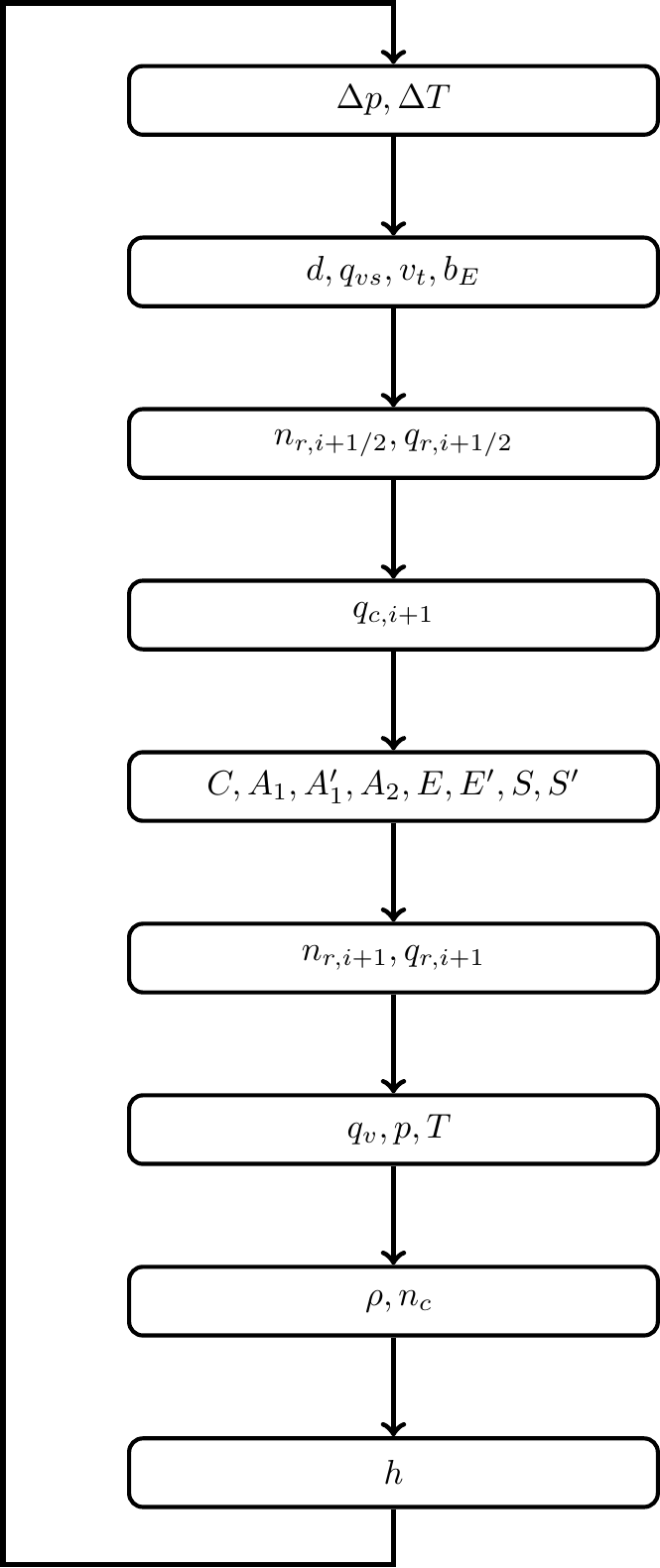}
	\caption{Flowchart of the numerical algorithm for the integration of the cloud model.}
	\label{Fig:flowchart}
\end{figure}

A flowchart of the numerical algorithm is presented in
Figure~\ref{Fig:flowchart}. 
%\section{List of constants}
%\label{appendixB:constants}
The physical constants and the model parameters are summarized in 
Tables~\ref{tab:appendix_physical_constants} and 
\ref{tab:appendix_model_parameters}, respectively.
Finally, we use the parameterization 
\begin{equation}
  \label{eq:appendixB_pliq}
  \begin{aligned}
    \log(p_{\mathrm{s}}(T)) & = 54.842763 - 6763.22/T -
    4.210\log(T) + 0.000367T\\ 
    &\quad + \tanh(0.0415(T-218.8))\\
    &\qquad\cdot\qty(53.878 -
    1331.22/T - 9.44523\log(T) +
    0.014025T)
  \end{aligned}
\end{equation}
for the saturation vapour pressure over a flat surface
of water from \citep{murphy_koop2005}.

\begin{table}
	\centering
	\begin{tabular}{ll}
		Constant & Description \\
		\hline
		$p_{\ast}=\SI{101325}{\pascal}$ & reference pressure\\
		$T_{\ast}=\SI{288}{\kelvin}$ & reference temperature\\
		$T_0=\SI{273.15}{\kelvin}$ & melting temperature \\
		$\rho_{\ast}=\SI{1.225}{\kilogram\per\cubic\meter}$ & reference air density\\
		$\gamma = \frac{g}{c_p}=\SI{0.00976}{\kelvin\per\meter}$ & dry adiabatic lapse rate\\
		$\rho_l=\SI{1000}{\kilogram\per\cubic\meter}$ & density of liquid water\\
		$R_a=\SI{287.05}{\joule\per\kilogram\per\kelvin}$ & specific gas constant, dry
		air\\
		$R_v=\SI{461.52}{\joule\per\kilogram\per\kelvin}$ & specific gas constant, water
		vapour\\
		$c_p=\SI{1005}{\joule\per\kilogram\per\kelvin}$ & specific heat capacity, dry
		air\\
		$g=\SI{9.81}{\meter\per\square\second}$ & acceleration due to gravity\\
		$L=\SI{2.53e6}{\joule\per\kilogram}$ & latent heat of vapourisation\\
		$\varepsilon=\frac{M_{\text{mol},v}}{M_{\text{mol},a}}=0.622$ &
		ratio
		of
		molar
		masses
		of
		water
		and  
		dry
		air\\
		$D_0=\SI{2.11e-5}{\square\meter\per\second}$ & diffusivity constant\\
	\end{tabular}
	\caption{Physical constants and reference quantities.}
	\label{tab:appendix_physical_constants}
\end{table}

\begin{table}
	\centering
	\begin{tabular}{ll}
		Parameter & Description \\
		\hline
		$\alpha=\SI{190.3}{\meter\per\second\raiseto{-\beta}\kilogram}$ & parameter for
		terminal
		velocity\\ 
		$\beta=\frac{4}{15}$ & parameter for terminal velocity\\
		$m_t=\SI{1.21e-5}{\kilogram}$ & parameter for terminal velocity\\
		% $c_n$ & parameter for number weighted terminal velocity\\
		% $c_q$ & parameter for mass weighted terminal velocity\\
		$k_1=\SI{0.0041}{\kilogram\per\second}$ & parameter for autoconversion\\
		$k_2=\SI{0.8}{\kilogram}$ & parameter for accretion\\
		$N_0=\SI{1000}{\per\cubic\meter}$ & parameter for activation\\
		$N_\infty$ & parameter for activation\\
		$m_0=\frac{4}{3}\pi\rho_l\qty(\SI{0.5}{\micro\meter})^3$ & parameter for activation\\
		$a_E= 0.78$ & parameter for evaporation\\
		$a_v= 0.78$ & parameter for ventilation\\
		$b_v= 0.308$ & parameter for ventilation\\
		$c_q= 1.84$ & parameter for effective fall velocity\\
		$c_n= 0.58$ & parameter for effective fall velocity\\
		$h$ & height of the box\\
	\end{tabular}
	\caption{Model parameters.}
	\label{tab:appendix_model_parameters}
\end{table}

\printbibliography

\end{document}